\documentclass[a4paper,11pt]{article}
\usepackage{amsfonts}

\usepackage{graphicx}
\graphicspath{{Figures/}}

\usepackage{amsmath}
\usepackage{amssymb}
\usepackage{latexsym}
\usepackage[colorlinks]{hyperref}
\usepackage{color}
\usepackage{float}
\usepackage{cite}
\usepackage{makeidx}
\usepackage{colortbl}
\usepackage{csquotes}
\usepackage[squaren]{SIunits}
\usepackage{physics}

\usepackage[textfont={footnotesize,sf},labelfont={color=blue,bf,sf},labelsep=endash]{caption}
\usepackage[position=top,labelfont={color=blue,bf,sf}]{subfig}
\usepackage{bm}

\newcommand{\rr}{\color{red}}

\usepackage[title]{appendix}

\begin{document}
\begin{titlepage}
\setcounter{page}{1}
\renewcommand{\thefootnote}{\fnsymbol{footnote}}

\begin{center}
{\Large \bf { {Implementing quantum Fourier transform  using three  qubits  
}}}

\vspace{5mm}

{
{\bf Mouhcine Yachi}$^{a}$,	{\bf Radouan Hab-arrih}$^{a}$ and {\bf Ahmed Jellal\footnote{\sf a.jellal@ucd.ac.ma}}$^{a,b}$
}

\vspace{5mm}
{$^{a}$\em Laboratory of Theoretical Physics,
	Faculty of Sciences, Choua\"ib Doukkali University},\\
{\em PO Box 20, 24000 El Jadida, Morocco}

{$^{b}$\em Canadian Quantum  Research Center,
	204-3002 32 Ave Vernon, \\ BC V1T 2L7,  Canada}

\vspace{30mm}
\begin{abstract}
     
	 {Using the circulant symmetry of a Hamiltonian describing three qubits, we realize the quantum Fourier transform.}
    This symmetry allows us to construct a set of eigenvectors independently on the magnitude of physical parameters {involved in the Hamiltonian} and as a result the entanglement will be  {maintained}. {The realization  will be leaned on trapped ions and the gate implementation} requires an adiabatic transition from each spin product state to Fourier modes. The fidelity was numerically calculated and the results show important values. Finally, {we discuss the acceleration of the gate by using the counter-driving field.}

    \vspace{50mm}
	
	\noindent PACS numbers:  03.65.Fd, 03.65.Ge, 03.65.Ud, 03.67.Hk
\\  
	\noindent Keywords: Three qubits, circulant symmetry,
	entanglement, quantum Fourier transform, adiabatic transition, counter-driving-field.
\end{abstract}
\end{center}
\end{titlepage}

\section{Introduction}

{Paul Benioff suggested a quantum mechanical model of the Turing machine \cite{Benioff1980} in 1980, launching a new field of study known as quantum computers (QCs).
Richard Feynman demonstrated in 1982 that QCs may be used to mimic complicated systems (living cells, city traffic, the human brain, and the universe, $\cdots$) \cite{Richard}. 
Seth Lloyd demonstrated four decades later that the basic units of a QC are an array of nuclear magnetic resonance spins \cite{Lyod}.
QC is a device that uses quantum mechanics to process data while maintaining quantum coherence \cite{ref1}.
As a result, QC can tackle the most difficult computational problems that even today's most powerful supercomputers cannot  solve \cite{shor}.
} 
As an example, Shor’s quantum algorithm (SQA) for factoring large numbers\cite{shoralg} is the most seminal motivation behind the development of QCs \cite{shor}. It is well-known in quantum computing that the physical implementation of SQA requires a gate of particular importance called quantum Fourier transform (QFT)\cite{qft}.
QFT is the quantum implementation of the discrete Fourier transform over the amplitudes of a wavefunction, {which acts on a vector $|x\rangle\in \mathbb{C}^{N} $ as
$|x \rangle\xrightarrow{\text{QFT}} |y\rangle=\frac{1}{\sqrt{N}}\sum_{j=0}^{N-1} e^{2\pi i x j/N}|j\rangle$} \cite{shor}. QFT is widely used as a key subroutine in several factoring algorithms like  for instance quantum amplitude estimation \cite{est1} and quantum counting \cite{count1}.  

{The circulant matrices have been thoroughly investigated in \cite{circ3,Gray2006}, and the set of invertible circulant matrices forms a group in terms of matrix multiplication.
	Geometry, linear coding, and graph theory, 
	 all use such matrices, one may see 
	\cite{Dzhelepov2011,Jiang2013, Muzychuk2004, Olson2014, Razpopov2015,Roth1990}.
Also for more than 47 years, vibration analysis has focused on systems with circulant  symmetry.
Many of these contributions were motivated by vibration
studies of general rotationally periodic systems \cite{Olson2014},  bladed
disks, planetary gear systems, rings,
circular plates, disk spindle systems, centrifugal
pendulum vibration absorbers, space antennae, and
microelectromechanical system frequency filters.
}

{The circulant Hamiltonians are addressed \cite{circ1} in terms of}
the physical implementation of logical QFT,  which {related} to the fact that the eigenspectrum of circulant matrix is spanned by the Fourier modes\cite{Eigen,circ3}. 
{Ivanov and Vitanov \cite{circ2} recently proposed a Hamiltonian based on two spins emerging in a magnetic field, creating Rabi oscillations, and obtained circulant symmetry by modifying the coupling strength of the spin-spin interaction. 
As a result, they demonstrated that the eigenvectors are independent of the magnitude of the physical parameters, implying entanglement protection, and the resulting system can subsequently be employed as a logical QFT. 
}
{Wu and his colleagues \cite{adiapass} presented two approaches for implementing quantum phase gates based on adiabatic passing and phase control of the driving fields.}
They studied 
the experimental feasibility, gate fidelity and
docoherence effect for both schemes. Moreover, 
the shortcuts to adiabaticity have become instrumental in preparing and driving internal and motional
states in atomic, molecular and solid-state physics,   {for a complete review we refer to} \cite{Adia}.

Motivated by the results developed in \cite{circ2}, we study a Hamiltonian describing three 
spins in a magnetic field, coupled via linear and non-linear interactions. The
last  
coupling generally arises when the interaction medium is non-linear \cite{NL,NL1,NL2} 
and remains an important ingredient  {in generating} a circulant Hamiltonian. 
{To build a logical QFT, we propose two schemes  based on the  physical parameter choices. The eigenvectors of our circulant Hamiltonian do not depend on the parameters, which protects entanglement during the gate implementation as long as the circulant symmetry is maintained. We use an energy offset Hamiltonian $ {H}_{0}(t)=\sum\limits_{j=1}^{3}\Delta_{j}(t){\sigma}_{j}^{z} $ to break the circulant symmetry during the transition.}
{We adjust} the physical parameters and the detuning $ \Delta_{j}(t) $ {to}
realize the circulant symmetry at the end of the transition $ t_{f} $. {We sinusoidally modulate   the
physical parameters in time to} show that it is possible to adiabatically obtain 
{a superposition of the quantum Fourier modes
from any initial state  with high fidelity.} As the adiabatic evolution is robust but  still limited by the non-adiabatic transition,  we  introduce  a counter-driving Hamiltonian 
${H}_{\sf CD}(t) $\cite{CD} to suppress these transitions. Under suitable conditions 
of the physical parameters, we determine the eigenvectors associated {with} the total Hamiltonian. These  allow us to combine our gate with the short-cut to adiabacity scheme in order  to accelerate the gate. 
{Based on the proposal described in \cite{circ2}, we suggest a way to}
physically implement the gate.

The outlines of our paper are summarized 
as follows. In {\color{red}{Sec}}. \ref{sec2}, {we propose a Hamiltonian describing three qubits with different interactions in addition to Rabi oscillations and show how to obtain the circulant symmetry.} The adiabatic transition technique is performed  to end up with the Fourier modes in {\color{red}{Sec}}. \ref{sec3}.  The physical implementation of the obtained  QFT gate will be discussed
in {\color{red}{Sec}}. \ref{sec4}. {The non-degeneracy of frequencies, the fidelities of our gate and the creation of entangled state will be numerically analyzed by sinusoidally varying the physical parameters} in {\color{red}{Sec}}. \ref{sec5}. We discuss the shortcut to adiabacity scheme combined with our gate 
in {\color{red}{Sec}}. \ref{sec6}. Finally, we close by  concluding our work.

\section{Theoretical model 
\label{sec2} }
\begin{figure}[htbphtbp]
	\centering 
	\includegraphics[width=9cm,height=5cm]{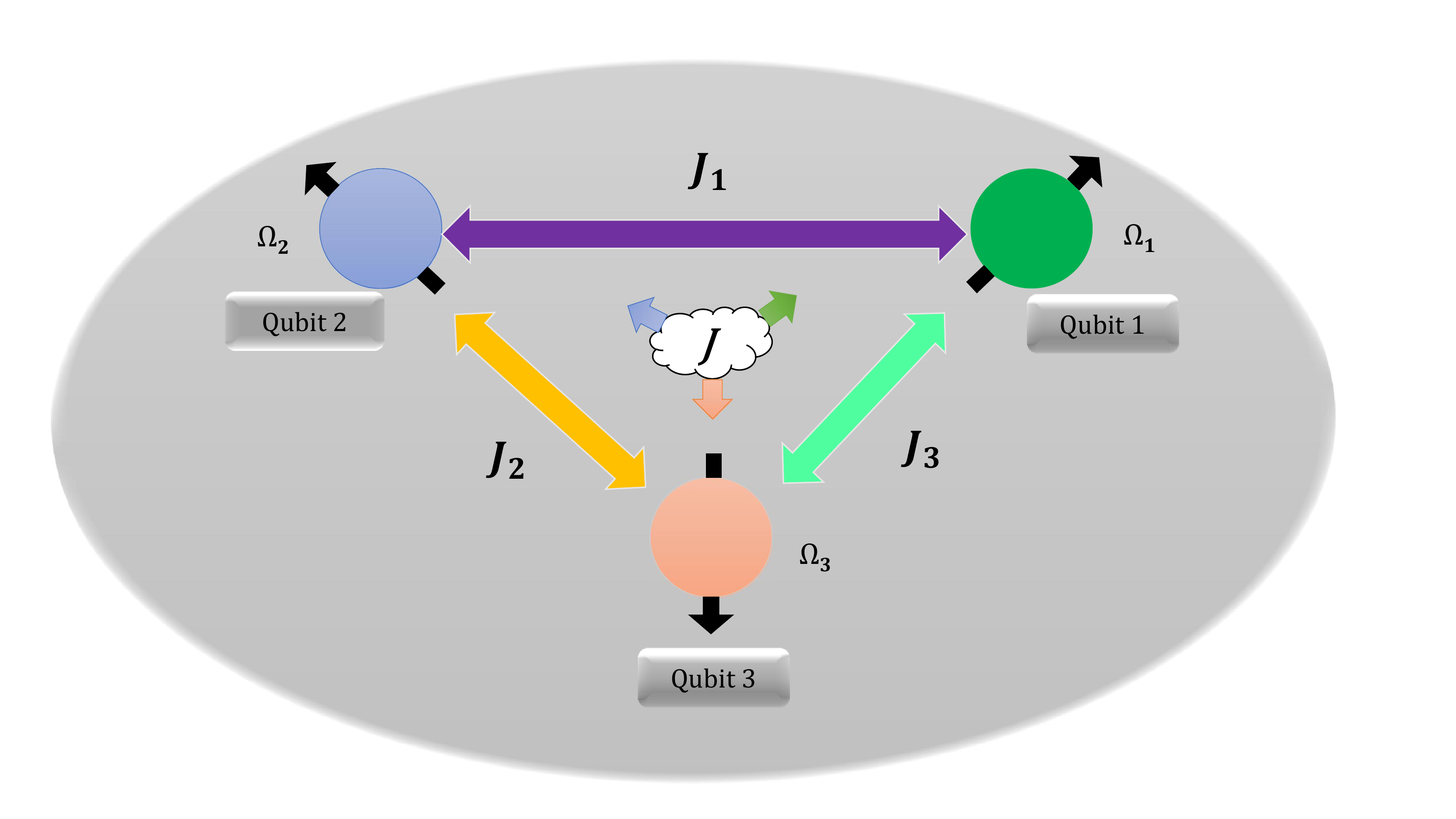}
	\captionof{figure}{(color online) The schematic presents the interactions {of strengths $J_j$} between three coupled qubits 
		{emerging} in  a magnetic field {with the Rabi frequencies $ \Omega_{j} $}.}
	\label{3qub}
\end{figure}

{We consider three spins that emerged in a magnetic field to achieve our task,} forming a three-qubit gate system
as presented in {\color{red}Figure \ref{3qub}}. 
\noindent

It can be described by a Hamiltonian involving two types of interaction, such as 
\begin{align}\label{1}
	\nonumber
{H}=&J_{1}(\sigma_{1}^{+} e^{-i\phi_{12}}+\sigma_{1}^{-} e^{i\phi_{12}})(\sigma_{2}^{+} e^{-i\phi_{21}}+\sigma_{2}^{-} e^{i\phi_{21}})+J_{2}(\sigma_{2}^{+} e^{-i\phi_{23}}+\sigma_{2}^{-} e^{i\phi_{23}})(\sigma_{3}^{+} e^{-i\phi_{32}}+\sigma_{3}^{-} e^{i\phi_{32}})\\ \nonumber 
&+J_{3}(\sigma_{1}^{+} e^{-i\phi_{13}}+\sigma_{1}^{-} e^{i\phi_{13}})(\sigma_{3}^{+} e^{-i\phi_{31}}+\sigma_{3}^{-} e^{i\phi_{31}})+\Omega_{1}(\sigma_{1}^{+}e^{i\theta_{1}}+\sigma_{1}^{-}e^{-i\theta_{1}})+\Omega_{2}(\sigma_{2}^{+}e^{i\theta_{2}}+\sigma_{2}^{-}e^{-i\theta_{2}})\\  
&+\Omega_{3}(\sigma_{3}^{+}e^{i\theta_{3}}+\sigma_{3}^{-}e^{-i\theta_{3}})
+J(\sigma_{1}^{+} e^{-i\phi_{1}}+\sigma_{1}^{-} e^{i\phi_{1}})(\sigma_{2}^{+} e^{-i\phi_{2}}+\sigma_{2}^{-} e^{i\phi_{2}})(\sigma_{3}^{+} e^{-i\phi_{3}}+\sigma_{3}^{-} e^{i\phi_{3}})
\end{align}
where $ \sigma_{j}^{+} =\ket{\uparrow_{j}} \bra{\downarrow_{j}}   $ and $ \sigma_{j}^{-} =\ket{\downarrow_{j}} \bra{ \uparrow_{j} }   $ stand for  spin flip operators, $ \ket{\uparrow_{j}} $ and $ \ket{\downarrow_{j}}$ being the qubit states of the $j^{th}$ spin, with  $ j = 1, 2,3$. The first term describes the interaction between spins $ 1 $ and $ 2 $ of  phases $ \phi_{12} $ and $ \phi_{21} $, {whereas the second term covers
the interaction
between spins 
$ 2 $ and $ 3 $ of phases  $ \phi_{23} $ and $ \phi_{32} $, and  the third term  describes the interaction between
spins $ 1 $ and $ 3 $ of phases $ \phi_{13} $ and $ \phi_{31} $}. The terms,
including
the Rabi frequencies $\Omega_{j}$, 
are the single-qubit transitions with  phases $ \theta_{j}$. The last term describes the coupling between  the three spins with phases $ \phi_{j} $ \cite{NL2}. {Note that, 	our Hamiltonian can be considered as a three-qubit generalization of the Ivanov and Vitanov model for two qubits \cite{circ2}, thanks to the trilinear interaction term.} 

It is convenient for our task to consider the matrix form of the Hamiltonian 
 \eqref{1} and then,  
in the basis $\mathcal{B}_{c}=\left\lbrace  \mid\downarrow\downarrow\downarrow\rangle ,\mid\downarrow\downarrow\uparrow\rangle,\mid\downarrow\uparrow\downarrow\rangle,\mid\downarrow\uparrow\uparrow\rangle,\mid\uparrow\downarrow\downarrow\rangle,\mid\uparrow\downarrow\uparrow\rangle,\mid\uparrow\uparrow\downarrow\rangle,\mid\uparrow\uparrow\uparrow\rangle \right\rbrace$, we have 
\begin{align}
{H}=\begin{pmatrix}
		0 & \Omega_{3}e^{i\theta_{3}} & \Omega_{2}e^{i\theta_{2}}  & J_{2}e^{-i\xi_{1}}  &\Omega_{1}e^{i\theta_{1}} & J_{3}e^{-i\xi_{2}} & J_{1}e^{-i\xi_{3}} &Je^{-i\xi_{7}}  \\
		\Omega_{3}e^{-i\theta_{3}}& 0 & J_{2}e^{-i\xi_{4}} & \Omega_{2}e^{i\theta_{2}} & J_{3}e^{-i\xi_{5}} & \Omega_{1}e^{i\theta_{1}} &Je^{-i\xi_{8}}  &J_{1}e^{-i\xi_{3}} \\
		\Omega_{2}e^{-i\theta_{2}}& J_{2}e^{i\xi_{4}} & 0 & \Omega_{3}e^{i\theta_{3}} & J_{1}e^{-i\xi_{6}} & Je^{-i\xi_{9}} & \Omega_{1}e^{i\theta_{1}} & J_{3}e^{-i\xi_{2}} \\
		J_{2}e^{i\xi_{1}}& \Omega_{2}e^{-i\theta_{2}} & \Omega_{3}e^{-i\theta_{3}} & 0 & Je^{-i\xi_{10}} &  J_{1}e^{-i\xi_{6}}& J_{3}e^{-i\xi_{5}} & \Omega_{1}e^{i\theta_{1}} \\
		\Omega_{1}e^{-i\theta_{1}}& J_{3}e^{i\xi_{5}} & J_{1}e^{i\xi_{6}} & Je^{i\xi_{10}} &  0& \Omega_{3}e^{i\theta_{3}} & \Omega_{2}e^{i\theta_{2}} & J_{2}e^{-i\xi_{1}} \\
		J_{3}e^{i\xi_{2}}& \Omega_{1}e^{-i\theta_{1}} & Je^{i\xi_{9}} & J_{1}e^{i\xi_{6}} &  \Omega_{3}e^{-i\theta_{3}}& 0 & J_{2}e^{-i\xi_{4}} & \Omega_{2}e^{i\theta_{2}} \\
		J_{1}e^{i\xi_{3}}& Je^{i\xi_{8}} &  \Omega_{1}e^{-i\theta_{1}}& J_{3}e^{i\xi_{5}} & \Omega_{2}e^{-i\theta_{2}} & J_{2}e^{i\xi_{4}} & 0 &\Omega_{3}e^{i\theta_{3}}  \\
		Je^{i\xi_{7}}& J_{1}e^{i\xi_{3}} &  J_{3}e^{i\xi_{2}}& \Omega_{1}e^{-i\theta_{1}} & J_{2}e^{i\xi_{1}} & \Omega_{2}e^{-i\theta_{2}} & \Omega_{3}e^{-i\theta_{3}} &0 
	\end{pmatrix} \label{2}
\end{align}
where  {the ten angles}
$\xi_{1}=\phi_{23}+\phi_{32}$, $\xi_{2}=\phi_{13}+\phi_{31}$, $\xi_{3}=\phi_{12}+\phi_{21}$, $ \xi_{4}=\phi_{23}-\phi_{32}, $
$\xi_{5}=\phi_{13}-\phi_{31}$,
$ \xi_{6}=\phi_{12}-\phi_{21},$
$  \xi_{7}=\phi_{1}+\phi_{2}+\phi_{3},$ $ \xi_{8}=\phi_{1}+\phi_{2}-\phi_{3},$ $\xi_{9}=\phi_{1}-\phi_{2}+\phi_{3},$ $\xi_{10}=\phi_{1}-\phi_{2}-\phi_{3} $ {have been introduced.}

In what follows, we are going  to find  conditions on the involved parameters  to end up with the Hamiltonian (\ref{2}) as a circulant matrix \cite{Eigen,circ3}. The benefit of {the}
circulant matrix is that its eigenvectors are the vectors of columns of the discrete quantum Fourier transform, and therefore they do not depend on the elements of the circulant matrix. The eigenvectors of our circulant matrix can be mapped {into} the spin basis $\mathcal{B}_{c}$ as
\begin{align}
	&\ket{\psi_{0}}=\dfrac{1}{2\sqrt{2}}(\ket{\downarrow\downarrow\downarrow}+\ket{\downarrow\downarrow\uparrow}+\ket{\downarrow\uparrow\downarrow}+\ket{\downarrow\uparrow\uparrow}+\ket{\uparrow\downarrow\downarrow}+\ket{\uparrow\downarrow\uparrow}+\ket{\uparrow\uparrow\downarrow}+\ket{\uparrow\uparrow\uparrow}) \label{3} \\
	&\ket{\psi_{1}}=\dfrac{1}{2\sqrt{2}}(\ket{\downarrow\downarrow\downarrow} +\omega\ket{\downarrow\downarrow\uparrow}+i\ket{\downarrow\uparrow\downarrow}+i\omega\ket{\downarrow\uparrow\uparrow}-\ket{\uparrow\downarrow\downarrow}-\omega\ket{\uparrow\downarrow\uparrow}-i\ket{\uparrow\uparrow\downarrow}-i\omega\ket{\uparrow\uparrow\uparrow}) \label{5}\\
&	\ket{\psi_{2}}=\dfrac{1}{2\sqrt{2}}(\ket{\downarrow\downarrow\downarrow} +i\ket{\downarrow\downarrow\uparrow}-\ket{\downarrow\uparrow\downarrow}-i\ket{\downarrow\uparrow\uparrow}+\ket{\uparrow\downarrow\downarrow}+i\ket{\uparrow\downarrow\uparrow}-\ket{\uparrow\uparrow\downarrow}-i\ket{\uparrow\uparrow\uparrow})\label{6} \\
&	\ket{\psi_{3}}=\dfrac{1}{2\sqrt{2}}(\ket{\downarrow\downarrow\downarrow} +i\omega\ket{\downarrow\downarrow\uparrow}-i\ket{\downarrow\uparrow\downarrow}+\omega\ket{\downarrow\uparrow\uparrow}-\ket{\uparrow\downarrow\downarrow}-i\omega\ket{\uparrow\downarrow\uparrow}+i\ket{\uparrow\uparrow\downarrow}-\omega\ket{\uparrow\uparrow\uparrow})\label{7}\\
&	\ket{\psi_{4}}=\dfrac{1}{2\sqrt{2}}(\ket{\downarrow\downarrow\downarrow} -\ket{\downarrow\downarrow\uparrow}+\ket{\downarrow\uparrow\downarrow}-\ket{\downarrow\uparrow\uparrow}+\ket{\uparrow\downarrow\downarrow}-\ket{\uparrow\downarrow\uparrow}+\ket{\uparrow\uparrow\downarrow}-\ket{\uparrow\uparrow\uparrow})\label{8}\\
&	\ket{\psi_{5}}=\dfrac{1}{2\sqrt{2}}(\ket{\downarrow\downarrow\downarrow}-\omega\ket{\downarrow\downarrow\uparrow}+i\ket{\downarrow\uparrow\downarrow}-i\omega\ket{\downarrow\uparrow\uparrow}-\ket{\uparrow\downarrow\downarrow}+\omega\ket{\uparrow\downarrow\uparrow}-i\ket{\uparrow\uparrow\downarrow}+i\omega\ket{\uparrow\uparrow\uparrow})\label{9}\\
&	\ket{\psi_{6}}=\dfrac{1}{2\sqrt{2}}(\ket{\downarrow\downarrow\downarrow}-i\ket{\downarrow\downarrow\uparrow}-\ket{\downarrow\uparrow\downarrow}+i\ket{\downarrow\uparrow\uparrow}+\ket{\uparrow\downarrow\downarrow}-i\ket{\uparrow\downarrow\uparrow}-\ket{\uparrow\uparrow\downarrow}+i\ket{\uparrow\uparrow\uparrow})\label{10}\\
&	\ket{\psi_{7}}=\dfrac{1}{2\sqrt{2}}\left(\ket{\downarrow\downarrow\downarrow}-i\omega\ket{\downarrow\downarrow\uparrow}-i\ket{\downarrow\uparrow\downarrow}-\omega\ket{\downarrow\uparrow\uparrow}-\ket{\uparrow\downarrow\downarrow}+i\omega\ket{\uparrow\downarrow\uparrow}+i\ket{\uparrow\uparrow\downarrow}+\omega\ket{\uparrow\uparrow\uparrow}\right)\label{11}
\end{align}
{and we have set}  the phase factor $ \omega = \exp(i\frac{\pi}{4}) $. {At this point, we show the possibilities that lead to the circulant symmetries. We end up with two scenarios when we require adequate requirements to be met by the physical parameters. Indeed,  the first configuration is}
%
\begin{align}
	&J_{1}=\Omega_{2}\\
	& J=J_{2}=J_{3}=\Omega_{3}\\
	& \Omega_{1}=0\\
	 &\theta_{2}=\theta_{3}=\phi_{32}=\phi_{3}=\phi_{21}=-\phi_{31}=\varphi\\
&\theta_{1}=\phi_{23}=\phi_{2}=\phi_{1}=\phi_{12}=\phi_{13}= 0
\end{align}
{which}
can be injected into \eqref{2} to 
get the first circulant Hamiltonian 
\begin{align}
			{H}_{\sf cir}^{(1)}=\begin{pmatrix}
				0&Je^{i\varphi}&J_{1}e^{i\varphi}&Je^{-i\varphi}&0&Je^{i\varphi}&J_{1}e^{-i\varphi}&Je^{-i\varphi}  \\
				Je^{-i\varphi}&0&Je^{i\varphi}&J_{1}e^{i\varphi}&Je^{-i\varphi}&0&Je^{i\varphi}&J_{1}e^{-i\varphi}\\ 
				J_{1}e^{-i\varphi}&Je^{-i\varphi}&0&Je^{i\varphi}&J_{1}e^{i\varphi}  &Je^{-i\varphi}&0&Je^{i\varphi} \\
				Je^{i\varphi}&J_{1}e^{-i\varphi} & Je^{-i\varphi} & 0 & Je^{i\varphi} & J_{1}e^{i\varphi} & Je^{-i\varphi} & 0 \\
				0	& Je^{i\varphi} & J_{1}e^{-i\varphi} & Je^{-i\varphi}& 0 & Je^{i\varphi} &  J_{1}e^{i\varphi}& Je^{-i\varphi} \\
				Je^{-i\varphi}	&0& Je^{i\varphi} & J_{1}e^{-i\varphi} & Je^{-i\varphi}& 0 & Je^{i\varphi} &  J_{1}e^{i\varphi}\\
				J_{1}e^{i\varphi}&Je^{-i\varphi}&0&Je^{i\varphi}&J_{1}e^{-i\varphi}&Je^{-i\varphi}&0&Je^{i\varphi}  \\
				Je^{i\varphi}&J_{1}e^{i\varphi}&Je^{-i\varphi}&0&Je^{i\varphi}&J_{1}e^{-i\varphi}& Je^{-i\varphi} & 0
			\end{pmatrix} \label{4}
\end{align}
{whereas the second configuration looks like }
\begin{align}
&	J_{1}=\Omega_{2}\\
& J=J_{2}=J_{3}=\Omega_{3}\\
& \theta_{2}=\theta_{3}=\phi_{32}=\phi_{3}=\phi_{21}=-\phi_{31}=\varphi\\
&\theta_{1}=\phi_{23}=\phi_{2}=\phi_{1}=\phi_{12}=\phi_{13}= 0
\end{align}
{with} $\Omega_{1}\neq 0$ and then, from \eqref{2} we obtain the second circulant Hamiltonian 
\begin{align}\label{12}
			{H}_{\sf cir}^{(2)}=\begin{pmatrix}
				0&Je^{i\varphi}&J_{1}e^{i\varphi}&Je^{-i\varphi}&\Omega_{1}&Je^{i\varphi}&J_{1}e^{-i\varphi}&Je^{-i\varphi}  \\
				Je^{-i\varphi}&0&Je^{i\varphi}&J_{1}e^{i\varphi}&Je^{-i\varphi}&\Omega_{1}&Je^{i\varphi}&J_{1}e^{-i\varphi}\\ 
				J_{1}e^{-i\varphi}&Je^{-i\varphi}&0&Je^{i\varphi}&J_{1}e^{i\varphi}  &Je^{-i\varphi}&\Omega_{1}&Je^{i\varphi} \\
				Je^{i\varphi}&J_{1}e^{-i\varphi} & Je^{-i\varphi} & 0 & Je^{i\varphi} & J_{1}e^{i\varphi} & Je^{-i\varphi} & \Omega_{1} \\
				\Omega_{1}	& Je^{i\varphi} & J_{1}e^{-i\varphi} & Je^{-i\varphi}& 0 & Je^{i\varphi} &  J_{1}e^{i\varphi}& Je^{-i\varphi} \\
				Je^{-i\varphi}	&\Omega_{1}& Je^{i\varphi} & J_{1}e^{-i\varphi} & Je^{-i\varphi}& 0 & Je^{i\varphi} &  J_{1}e^{i\varphi}\\
				J_{1}e^{i\varphi}&Je^{-i\varphi}&\Omega_{1}&Je^{i\varphi}&J_{1}e^{-i\varphi}&Je^{-i\varphi}&0&Je^{i\varphi}  \\
				Je^{i\varphi}&J_{1}e^{i\varphi}&Je^{-i\varphi}&\Omega_{1}&Je^{i\varphi}&J_{1}e^{-i\varphi}& Je^{-i\varphi} & 0
			\end{pmatrix} .
\end{align}
{At this level, we underline that the presence of the trilinear interaction is necessary to realize the QFT gate. This type of interaction is required to achieve such circulant symmetry. Generalizing to more than three qubits is not trivial because the Hamiltonian will demand multilinear interactions. As a result, discussing the adiabatic transition and shortcut to adiabacity will be difficult mathematically.}
In the forthcoming analysis, we focus only on one of the above Hamiltonians,  {let say 
$ {H}_{\sf cir}^{(1)} $,} and investigate its basic features. 

\section{Adiabatic transition to Fourier modes \label{sec3} }

{Some controls can be used to create an adiabatic transition to Fourier modes.
	Two of these will be discussed here: the energy offset and Rabi frequencies. }


\subsection{Controlling by energy offset}
To realize an adiabatic evolution to the circulant Hamiltonian states (Fourier modes), we add an energy offset  $ {H}_{0}(t) $
\begin{equation}\label{hdet}
{H}_{0}(t)= \Delta_{1}(t){\sigma}_{1}^{z}+\Delta_{2}(t){\sigma}_{2}^{z}+\Delta_{3}(t){\sigma}_{3}^{z} 
\end{equation}
where  the time-dependent detuning $ \Delta_{j}(t) $ of $ j^{th} $ spin are necessary to control the adiabatic transition from computational spin states to quantum Fourier states (\ref{3}-\ref{11}). Consequently, we have now the Hamiltonian
\begin{equation}
		{H}(t)={H}_{0}(t)+{H}_{\sf cir}^{(1)}(t).
			\label{Of} 	
\end{equation}
{Remember that in the adiabatic limit, the system is always in the same  eigenstates of ${H}(t) $ \cite{circ2}.
The eigenstates of $ H(t) $ will be those of $ H_0 (t) $ at $ t_i $, and the dynamics will drive them to the Fourier modes at $ t_f $ if the couplings and detunings have a specific time dependency.
Therefore, adiabatic evolution translates each computational spin state to a Fourier mode, resulting in a single interaction step that generates the QFT.
The non-degeneracy between the eigen-frequencies of ${H}(t) $ must be bigger at any time than the non-adiabatic coupling between each pair of $ H(t) $ eigenstates $ \ket{\lambda_{\pm}} $, $ \ket{\delta_{\pm}} $, $ \ket{\mu_{\pm}} $, and $ \ket{\gamma_{\pm}} $ of  $ {H}(t) $, according to adiabatic evolution.}
Otherwise, we have {the conditions}
\begin{align} 
	&|\mu_{\pm}(t)-\lambda_{\pm}(t)| \gg|\bra{\partial_{t}\mu_{\pm}(t)}\ket{\lambda_{\pm}(t)}|\\
	&
	|\lambda_{+}(t)-\lambda_{-}(t)| \gg |\bra{\partial_{t}\lambda_{+}(t)}\ket{\lambda_{-}(t)}|\\ 
&	|\lambda_{\pm}(t)-\delta_{\pm}(t)| \gg |\bra{\partial_{t}\lambda_{\pm}(t)}\ket{\delta_{\pm}(t)}|\\
&
	|\delta_{+}(t)-\delta_{-}(t)| \gg |\bra{\partial_{t}\delta_{+}(t)}\ket{\delta_{-}(t)}|\\ 
	& |\delta_{\pm}(t)-\gamma_{\pm}(t)| \gg |\bra{\partial_{t}\delta_{\pm}(t)}\ket{\gamma_{\pm}(t)}|\\
	& 
	|\gamma_{+}(t)-\gamma_{-}(t)| \gg |\bra{\partial_{t}\gamma_{+}(t)}\ket{\gamma_{-}(t)}|. 
\end{align}

{Let us choose $ \varphi =\frac{\pi}{2} $ to simplify our problem, and then we prove in {\color{red}Appendix} \ref{appendixA}  that the eigenvalues $ \lambda_{\pm}, \delta_{\pm}, \mu_{\pm}, \gamma_{\pm} $  of the Hamiltonian ${H}(t) $ \eqref{Of} are provided by (\ref{a1}-\ref{a4}).}
%
Now, it is {worthwhile} to mention that the circulant symmetry is broken. We suppose that the system is initially prepared in the computational product states $ \ket{\psi_{s_{1}s_{2}s_{3}}} = \ket{s_{1}s_{2}s_{3}} $ $ (s_{j}=\downarrow_{j},\uparrow_{j}) $, which are eigenstates of the Hamiltonian $ {H}_{0}(t) $. As a result, the initial parameters should verify {the conditions}
\begin{eqnarray}
	\Delta_{1,2,3}(t_{i})\gg J(t_{i}),\:J_{1}(t_{i})
\end{eqnarray}
and then   ${H}(t)$ goes to ${H}_{0}(t) $. Consequently   the eigenvalues become 
\begin{align}
	&		\lambda_{\pm}(t_{i})=\pm\left [\Delta_{1}(t_{i})+\Delta_{2}(t_{i})+\Delta_{3}(t_{i})\right]\\
		&	\delta_{\pm}(t_{i})=\pm \left [\Delta_{1}(t_{i})+\Delta_{2}(t_{i})-\Delta_{3}(t_{i})\right]\\
&			\mu_{\pm}(t_{i})=\pm \left [\Delta_{1}(t_{i})-\Delta_{2}(t_{i})+\Delta_{3}(t_{i})\right]\\
	&		\gamma_{\pm}(t_{i})=\pm \left [\Delta_{1}(t_{i})-\Delta_{2}(t_{i})-\Delta_{3}(t_{i})\right]		
\end{align}
{and the $ {H}(t) $ eigenvectors are exactly} the computational spin states, i.e. $ \ket{\psi(t_{i})} = \ket{s_{1}s_{2}s_{3}} $,
\begin{align}
&	\ket{\lambda_{+}}=\ket{\downarrow\downarrow\downarrow}, \quad \ket{\lambda_{-}}=\ket{\uparrow\uparrow\uparrow} \\
& \ket{\delta_{+}}=\ket{\downarrow\downarrow\uparrow}, \qquad \ket{\delta_{-}}=\ket{\uparrow\uparrow\downarrow} \\ 
&\ket{\mu_{+}}=\ket{\downarrow\uparrow\downarrow}, \qquad \ket{\mu_{-}}=\ket{\uparrow\downarrow\uparrow} \\
&\ket{\gamma_{+}}=\ket{\downarrow\uparrow\uparrow}, \qquad\ket{\gamma_{-}}=\ket{\uparrow\downarrow\downarrow}.
\end{align}
{To avoid degeneracy, the condition} $ \Delta_{i}(t_{i})\neq \Delta_{j}(t_{i})$ with $i,j = 1,2,3$, and we can have equidistant eigen-frequencies by requiring 
$ \Delta_{1}(t_{i}) = 2 \Delta_{2}(t_{i}) =
4\Delta_{3}(t_{i}) $. 
Furthermore, to obtain the Fourier modes at {the} final time $t_f$ of transition, {the coupling} parameters together with detunings should verify the conditions  
\begin{eqnarray}
	\Delta_{1,2,3}(t_{f})\ll J(t_{f}),\:J_{1}(t_{f}).
\end{eqnarray}
{As a result, the total Hamiltonian evolves to a circulant one, i.e. $ {H}(t)\rightarrow {H}_{\sf cir}^{(1)}(t) $,}
 and {its}   eigenspectrum becomes that of $ {H}_{\sf cir}^{(1)}(t) $, {as shown below}
\begin{align}
&	\ket{\lambda_{+}}=\ket{\psi_{0}},\qquad\ket{\lambda_{-}}=\ket{\psi_{7}}\\
&\ket{\delta_{+}}=\ket{\psi_{1}},\qquad\ket{\delta_{-}}=\ket{\psi_{6}}  \\ 
	&\ket{\mu_{+}}=\ket{\psi_{2}},\qquad\ket{\mu_{-}}=\ket{\psi_{5}} \\
	&\ket{\gamma_{+}}=\ket{\psi_{3}},\qquad\ket{\gamma_{-}}=\ket{\psi_{4}}.
\end{align}
Additionally, the realization of the QFT relies on the adiabatic following of each of the instantaneous eigenvectors
{\begin{eqnarray}
			\ket{\downarrow\downarrow\downarrow}\:&\longrightarrow&\: e^{i\alpha_{1}}\:\ket{\psi_{0}}\\
			\ket{\downarrow\downarrow\uparrow}\:&\longrightarrow&\: e^{i(\alpha_{2}-\frac{\pi}{2})}\:\ket{\psi_{1}}	\\
			\ket{\downarrow\uparrow\downarrow}\:&\longrightarrow&\: e^{i\alpha_{3}}\:\ket{\psi_{2}}\\
			\ket{\downarrow\uparrow\uparrow}\:&\longrightarrow&\: e^{i\alpha_{4}}\:\ket{\psi_{3}}\\
			\ket{\uparrow\downarrow\downarrow}\:&\longrightarrow&\: e^{-i\alpha_{4}}\:\ket{\psi_{4}}	\\
			\ket{\uparrow\downarrow\uparrow}\:&\longrightarrow&\: e^{-i\alpha_{3}}\:\ket{\psi_{5}}	\\
			\ket{\uparrow\uparrow\downarrow}\:&\longrightarrow&\: e^{-i\alpha_{2}}\:\ket{\psi_{6}}\\
			\ket{\uparrow\uparrow\uparrow}\:&\longrightarrow&\: e^{-i\alpha_{1}}\:\ket{\psi_{7}}
\end{eqnarray}
{and} the global adiabatic phases {$\alpha_{j}$ appear} due to the adiabatic
evolution \cite{circ2,AdiaEvolution,berryphase,Adia}
\begin{align}
&	\alpha_{1}= \int_{t_{i}}^{t_{f}} \lambda_{+}(t) \, \mathrm{d}t,\qquad\alpha_{2}= \int_{t_{i}}^{t_{f}} \delta_{+}(t) \, \mathrm{d}t,\qquad\alpha_{3}=  \int_{t_{i}}^{t_{f}} \mu_{+}(t) \, \mathrm{d}t ,\qquad\alpha_{4}=  \int_{t_{i}}^{t_{f}} \gamma_{+}(t) \, \mathrm{d}t
\end{align}
{with the relations}
\begin{align}
\lambda_{-}(t)=-\lambda_{+}(t),\qquad \delta_{-}(t)=-\delta_{+}(t),\qquad \mu_{-}(t)=-\mu_{+}(t), \qquad \gamma_{-}(t)=-\gamma_{+}(t).
\end{align}
{After} a specific tuning  of the  detuning $\Delta_{j}(t)$, $\alpha_j$  reduce to
\begin{equation}\label{phases}
\alpha_{1}=2p\pi,\qquad \alpha_{2}=2m\pi,\qquad \alpha_{3}=2n\pi,\qquad \alpha_{4}=2k\pi
\end{equation}
with $ p, m, n, k $ are four integer numbers.
This choice leads to {the realization of} the following unitary quantum gate
\begin{align}
	\mathcal{G}=\dfrac{1}{2\sqrt{2}}\begin{pmatrix}
		1&-i&1&1&1&1&1&1 \\
		1&-i\omega&i&i\omega&-1&-\omega&-i&-i\omega  \\
		1&1&-1&-i&1&i&-1&-i  \\
		1&\omega&-i&\omega&-1&-i\omega&i&-\omega \\
		1&i&1&-1&1&-1&1&-1 \\
		1&i\omega&i&-i\omega&-1&\omega&-i&i\omega \\
		1&-1&-1&i&1&-i&-1&i \\
		1&-\omega&-i&-\omega&-1&i\omega&i&\omega
	\end{pmatrix}. \label{G}
\end{align}
As a result, one can show that
up to 
an additional phase $ -\frac{\pi}{2} $ the determinant of $ \mathcal{G}$ is equal {to} one, i.e.  $ \det(\mathcal{G})  =1$, which is necessary for adiabatic evolution. {Therefore, it is worthwhile} to note that the gate $ \mathcal{G} $ is a QFT one. 

\subsection{Controlling by Rabi frequencies}

Now we {will} discuss how to control the  Rabi frequencies $ \Omega_{2}(t) $  and $ \Omega_{3}(t) $ by considering  the Hamiltonian ${H}^{(1)}(t) $ \eqref{588} in {\color{red}Appendix} \ref{appendixB} without using the energy offset $ {H}_{0}(t) $ \eqref{hdet}. To achieve this goal, we drive 
$ \Omega_{2}(t) $ and $ \Omega_{3}(t) $ in a way that they {become} equal to the couplings $ J_{1} $ and $ J $, respectively.
{In what follows,} we summarize  the process of control in three steps.
\begin{itemize}
\item {\bf Initially:} Let us  assume that $ \Omega_{2}(t_{i})\gg J_{1} $  and $ \Omega_{3}(t_{i})\gg J $, then  the eigenvectors of ${H}^{(1)}(t) $ \eqref{588} are the rotating computational spin states
$ \ket{\psi({t_i})}=\ket{s'_{1}s'_{2}s'_{3}}$ ($s'_{j}=\pm_{j})$, 
such as
\begin{align}
&\ket{\pm_{1}}=\dfrac{1}{\sqrt{2}}\left(\ket{\downarrow_{1}}\pm\ket{\uparrow_{1}}\right)
\nonumber\\
& \ket{\pm_{2}}=\dfrac{1}{\sqrt{2}}\left(e^{i\varphi}\ket{\downarrow_{2}}\pm \: \ket{\uparrow_{2}}\right)\\
& \ket{\pm_{3}}=\dfrac{1}{\sqrt{2}}\left(\ket{\downarrow_{3}}\pm\ket{\uparrow_{3}}\right).
\nonumber
\end{align}
\item {\bf Transition:} The adiabatic transition from the initial eigenvectors to Fourier modes is given by the mappings 
\begin{eqnarray} \nonumber
			\ket{---}\:&\longrightarrow&\: e^{-i\beta_{0}}\:\ket{\psi_{0}} \\ \nonumber
			\ket{--+}\:&\longrightarrow&\: e^{-i\beta_{1}}\:\ket{\psi_{1}} 	\\ \nonumber
			\ket{-+-}\:&\longrightarrow&\: e^{-i\beta_{2}}\:\ket{\psi_{2}} \\ 
			\ket{-++}\:&\longrightarrow&\: e^{-i\beta_{3}}\:\ket{\psi_{3}}  \\ \nonumber
			\ket{+--}\:&\longrightarrow&\: e^{-i\beta_{4}}\:\ket{\psi_{4}} 	\\ \nonumber
			\ket{+-+}\:&\longrightarrow&\: e^{-i\beta_{5}}\:\ket{\psi_{5}} 	\\ \nonumber
			\ket{++-}\:&\longrightarrow&\: e^{-i\beta_{6}}\:\ket{\psi_{6}} \\ \nonumber
			\ket{+++}\:&\longrightarrow&\: e^{-i\beta_{7}}\:\ket{\psi_{7}} \label{41}
\end{eqnarray}
where the   adiabatic phases $ \beta_{i} $ read as
\begin{align}
\beta_{i}= \int_{t_{i}}^{t_{f}} \Lambda_{i}(t)\ dt, \qquad i=1,\cdots, 7 
\end{align} 
and {{\color{red}Appendix} \ref{appendixB} shows}
the eigenfrequencies $\Lambda_{i}(t)$ (\ref{bb8}-\ref{bb1}) of ${H}^{(1)}(t)$ \eqref{588}.

\item {\bf Finally:} We adiabatically decrease $ \Omega_{2}(t) $ together with $ \Omega_{3}(t) $ to end up with 
\begin{equation}
\Omega_{2}(t_{f})=J_{1}, \qquad  \Omega_{3}(t_{f})=J.
\end{equation}
As a result, the circulant symmetry will be established, and the Fourier modes will be derived as well.
\end{itemize}
At this stage, we mention that in the case of the gate realization based on the energy offset, the exact eigenvectors of the Hamiltonian \eqref{Of} can not be {obtained exactly}. However, under the {consideration made here,} the derivation of  eigenvectors can be achieved, see {\color{red}Appendix} \ref{appendixB}. Thus,  to accelerate the gate, we combine the gate scheme with the short-cut to adiabaticity.

\section{Gate implementation \label{sec4}}

To give a physical  implementation of our system, we generalize the process proposed by  Ivanov and  Vitanov
  \cite{circ2} 
to {the} three qubit case {with trilinear coupling}. Indeed, to realize the circulant Hamiltonian, we proceed with trapped ions \cite{Trepii4,Trepii5, Trepii66}. A crystal with $3N$  ions with mass $M$ is considered,  the trap axis are $ z $ and $ x $ with the frequencies $\Omega_{z}$ and $\Omega_{x}$, respectively. 
Each ion of the crystal is a qubit described by two typical levels: $ \ket{\uparrow},  \ket{\downarrow} $ with an energy gap  $ \omega_{0} $. Moreover, the virtual excitations generated between two coupled ions  undergo  small radial vibrations around the equilibrium positions of the ions. Then,
as an illustration, we {present} our physical implementation, 
depicted in {\color{red}Figure} \ref{3qubimple}.

\begin{figure}[htbphtbp]
 	\centering 
 	\includegraphics[width=9cm,height=6cm]{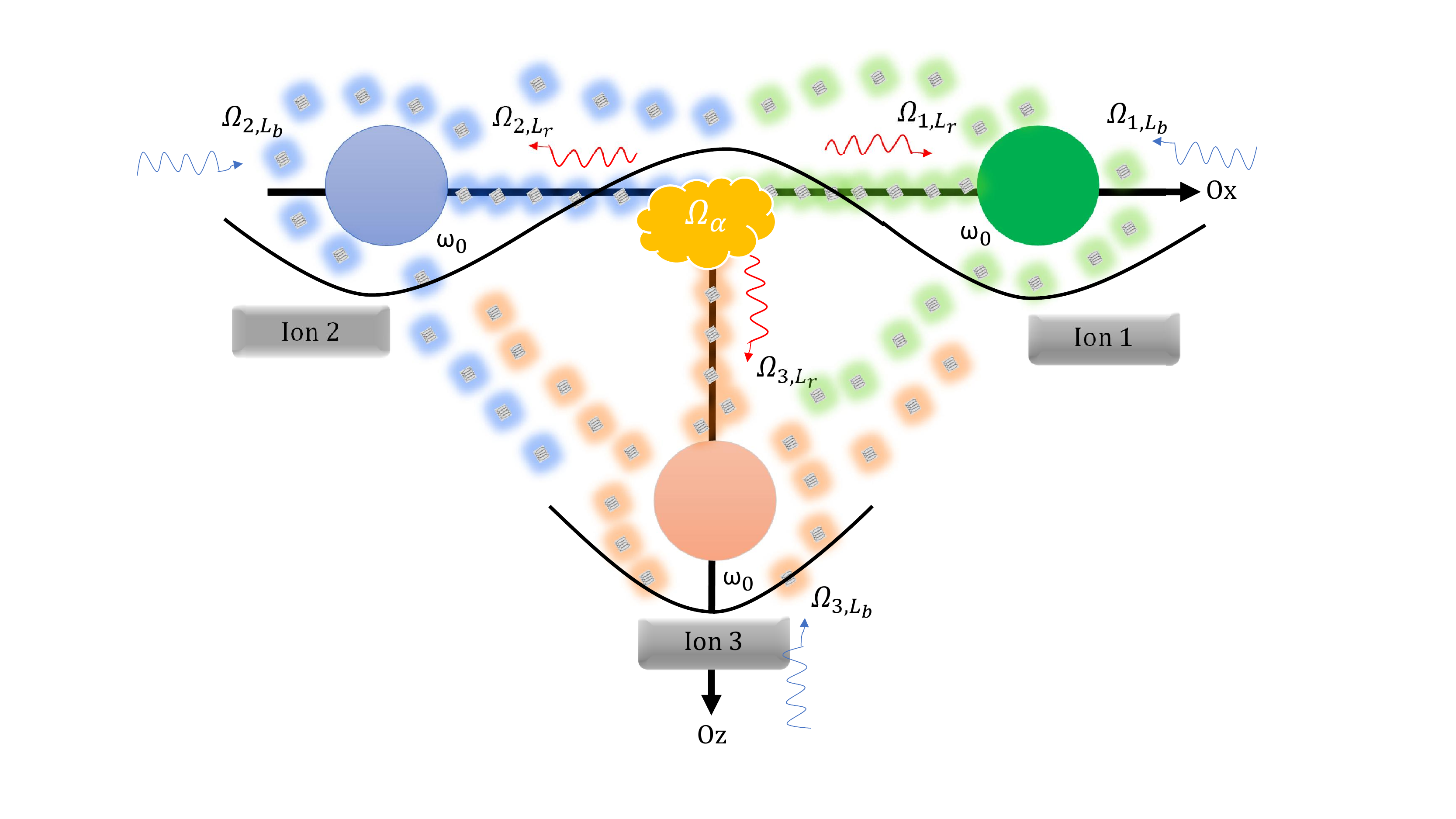}
 	\captionof{figure}{(color online) A proposal {for} the physical implementation of the interaction between three coupled trapped ions with laser frequencies $ \Omega_{j,L_{r}}= \omega_{0}-\nu-\Delta_{j}(t) $ and $ \Omega_{j,L_{b}}= \omega_{0}+\nu-\Delta_{j}(t) $ that generates  a spin-dependent force at the frequency $ \nu $.}
 	\label{3qubimple}
 \end{figure}
 
As for  our  Hamiltonian  (\ref{1}), the implementation of kinetic terms together with bilinear couplings is perfectly discussed in \cite{circ2}. Regarding    the implementation of the trilinear coupling, we suggest that it 
can be seen as a coupling between two coupled ions with an extra third ion \cite{NL2}. To be clear, 
 the small radial vibrations around the equilibrium positions between two coupled ions are displayed  by a set of collective vibrational modes with the Hamiltonians \cite{NL,paultrap}
\begin{equation}
{H}_{\sf ph1} = \sum_{n} \Omega_{n} \widehat{a}_{n}^{\dagger} \widehat{a}_{n},\qquad {H}_{\sf ph2} = \sum_{n} \Omega_{n} \widehat{b}_{n}^{\dagger} \widehat{b}_{n},\qquad  {H}_{\sf ph3} = \sum_{n} \Omega_{n} \widehat{c}_{n}^{\dagger} \widehat{c}_{n} 
\end{equation}
whereas the internal energy is 
\begin{equation}
{H}_{\sf q} =\frac{1}{2} \sum_{j} \omega_{0} \sigma_{j}^{z}.
\end{equation}
  and the free Hamiltonian {then} becomes 
\begin{equation}
{H}_{0}={H}_{\sf q}+\sum_{j} {H}_{\sf phj}.
\end{equation}
As claimed above, generating the trilinear coupling will be {dependent} on the interaction between two coupled ions {and} a third one. {For instance, this can be achieved}    by using two pairs of
 noncopropagating laser beams along the radial directions {of 
  frequencies 
\begin{align}
&\Omega_{j,L_{r}} = \omega_0-\nu-\Delta_j(t)\\  &\Omega_{j,L_{b}} =\omega_0+\nu-\Delta_j(t)
\end{align}  
 that} generate  a spin-dependent force at frequency $\nu$.

Besides, using the weak coupling assumption, one can legitimately apply the optical rotating-wave  {approximation and,  as a result}, 
the interacting part of our system reads now as  
\begin{align}\nonumber
{H}_{\sf I}=&\sum_{j}\Delta_{j}\sigma_{j}^{z} + \Omega_{x}  e^{ik\widehat{x}_{1}} \cos(\nu t)\left(\sigma_{1}^{+}e^{i\phi_{1}}+\sigma_{1}^{-}e^{-i\phi_{1}}\right)
+\Omega_{x}  e^{ik\widehat{x}_{2}} \cos(\nu t)\left(\sigma_{2}^{+}e^{i\phi_{2}}+\sigma_{2}^{-}e^{-i\phi_{2}}\right) \\ 
&+\Omega_{z}  e^{ik\widehat{z}_{3}} \cos(\nu t)\left(\sigma_{3}^{+}e^{i\phi_{3}}+\sigma_{3}^{-}e^{-i\phi_{3}}\right)
+\sum_{j}\Omega_{j}\left(\sigma_{j}^{+}e^{i\theta_{j}}+\sigma_{j}^{-}e^{-i\theta_{j}}\right)\\ \nonumber
&+\Omega_{\alpha} e^{ik\widehat{z}_{3}} \sin(\nu t)\left(\sigma_{1}^{+}e^{i{\varphi_{1}}}+\sigma_{1}^{-}e^{-i{\varphi_{1}}}\right)
\left(\sigma_{2}^{+}e^{i{\varphi_{2}}}+\sigma_{2}^{-}e^{-i{\varphi_{2}}}\right)
\end{align} 
where $\Omega_{x}, \Omega_{z}, \Omega_{j},  \Omega_{\alpha}$ 
are the Rabi frequencies, 
$ \phi_{j}, \theta_{j}$ are the laser phases, $ \varphi_{1,2} $ are the
phases resulted from trilinear interaction}, and the spatial arguments 
\begin{eqnarray}
		k\widehat{x}_{1}&=& \sum_{n} \eta_{1,n}(\widehat{a}_{n}^{\dagger}e^{i\Omega_{n}t}+\widehat{a}_{n}e^{-i\Omega_{n}t})\\
		k\widehat{x}_{2}&=& \sum_{n} \eta_{2,n}(\widehat{b}_{n}^{\dagger}e^{i\Omega_{n}t}+\widehat{b}_{n}e^{-i\Omega_{n}t})\\
		k\widehat{z}_{3}&=& \sum_{n} \eta_{3,n}(\widehat{c}_{n}^{\dagger}e^{i\Omega_{n}t}+\widehat{c}_{n}e^{-i\Omega_{n}t})
\end{eqnarray}
 {are} involving  the Lamb-Dicke parameters 
\begin{align}
\eta_{j,n}=b_{j,n}k\sqrt{\hbar/2M\Omega_{n}}
\end{align}
with   $ b_{j,n} $ are
the normal mode transformation matrix for the $j^{th}$ ion. {Hence, since} the dimensionless parameter $ \eta_{j,n} $ is small,  we can make the Lamb-Dicke approximation,  $ \Delta k \langle\widehat{x}_{1}\rangle \ll 1 $, $ \Delta k \langle\widehat{x}_{2}\rangle \ll 1 $, $ \Delta k \langle\widehat{z}_{3}\rangle \ll 1 $, to 
 end up with 
\begin{align}\nonumber
	{H}_{\sf I}=&\sum_{j}\Delta_{j}\sigma_{j}^{z} +\sum_{n}J_{1,n} \cos(\nu t)\left(\sigma_{1}^{+}e^{i\phi_{1}}+\sigma_{1}^{-}e^{-i\phi_{1}}\right)\left(\widehat{a}_{n}^{\dagger}e^{i\Omega_{n}t}+\widehat{a}_{n}e^{-i\Omega_{n}t}\right)\nonumber
	\\
	&
    +\sum_{j}\Omega_{j}\left(\sigma_{j}^{+}e^{i\theta_{j}}+\sigma_{j}^{-}e^{-i\theta_{j}}\right)
    +\sum_{n}J_{2,n} \cos(\nu t)\left(\sigma_{2}^{+}e^{i\phi_{2}}+\sigma_{2}^{-}e^{-i\phi_{2}}\right)\left(\widehat{b}_{n}^{\dagger}e^{i\Omega_{n}t}+\widehat{b}_{n}e^{-i\Omega_{n}t}\right)
\nonumber \\
&
	+\sum_{n}J_{3,n} \cos(\nu t)\left(\sigma_{3}^{+}e^{i\phi_{3}}+\sigma_{3}^{-}e^{-i\phi_{3}}\right) 
	\left(\widehat{c}_{n}^{\dagger}e^{i\Omega_{n}t}+\widehat{c}_{n}e^{-i\Omega_{n}t}\right) 
\\	
	&+\sum_{n}h_{n}\sin(\nu t)
	\left(\sigma_{1}^{+}e^{i{\varphi_{1}}}+\sigma_{1}^{-}e^{-i{\varphi_{1}}}\right)
	\left(\sigma_{2}^{+}e^{i{\varphi_{2}}}+\sigma_{2}^{-}e^{-i{\varphi_{2}}}\right)
	 \left(\widehat{c}_{n}^{\dagger}e^{i\Omega_{n}t}+\widehat{c}_{n}e^{-i\Omega_{n}t} \right)\nonumber 
\end{align}  
where $ J_{1,n} =\eta_{1,n}  \Omega_{x} $, $ J_{2,n} =\eta_{2,n}  \Omega_{x} $ and $ J_{3,n} =\eta_{3,n}  \Omega_{z} $ are the spin-phonon coupling and $ h_{n} =\eta_{3,n}\Omega_{\alpha}   $ is the  trilinear  coupling. Moreover, during a slow dynamics, the beat-note frequency $\nu$ isn't resonant  with any radial vibration mode, i.e.  $ |\Omega_{n}-\nu|\gg J_{j,n},h_{n} $. Additionally, {if} the phonons are virtually excited, then  they should be eliminated from the dynamics,  {and as {a} consequence}, the spin states  in different sites become coupled to each other.  For {the} different three sites  $ j^{th} $, $ p^{th} $ and $ q^{th}$, we have 
\begin{align}\nonumber
{H}_{\sf I}=&\Delta_{j}\sigma_{j}^{z}+
	\Delta_{p}\sigma_{p}^{z}+\Delta_{q}\sigma_{q}^{z}+\Omega_{j}\left(\sigma_{j}^{+}e^{i\theta_{j}}+\sigma_{j}^{-}e^{-i\theta_{j}}\right)+\Omega_{p}\left(\sigma_{p}^{+}e^{i\theta_{p}}+\sigma_{p}^{-}e^{-i\theta_{p}}\right) \\
	&+\Omega_{q}\left(\sigma_{q}^{+}e^{i\theta_{q}}+\sigma_{q}^{-}e^{-i\theta_{q}}\right)
	+J_{1}\left(\sigma_{j}^{+}e^{i\phi_{j}}+\sigma_{j}^{-}e^{-i\phi_{j}}\right)\left(\sigma_{p}^{+}e^{i\phi_{p}}+\sigma_{p}^{-}e^{-i\phi_{p}}\right) \\ \nonumber
	&+J_{2}\left(\sigma_{p}^{+}e^{i\phi_{p}}+\sigma_{p}^{-}e^{-i\phi_{p}}\right)\left(\sigma_{q}^{+}e^{i\phi_{q}}+\sigma_{q}^{-}e^{-i\phi_{q}}\right) 
	+J_{3}\left(\sigma_{j}^{+}e^{i\phi_{j}}+\sigma_{j}^{-}e^{-i\phi_{j}}\right)\left(\sigma_{q}^{+}e^{i\phi_{q}}+\sigma_{q}^{-}e^{-i\phi_{q}}\right)\\ \nonumber
	&+J\left(\sigma_{j}^{+}e^{i{\varphi_{j}}}+\sigma_{j}^{-}e^{-i{\varphi_{j}}}\right)\left(\sigma_{p}^{+}e^{i{\varphi_{p}}}+\sigma_{p}^{-}e^{-i{\varphi_{p}}}\right)\left(\sigma_{q}^{+}e^{i{\varphi_{q}}}+\sigma_{q}^{-}e^{-i{\varphi_{q}}}\right)
\end{align}
where the couplings between two ions are given by
\begin{align}
J_{1}=\sum_{n}J_{j,n}J_{p,n}\frac{1}{\nu^{2}-\Omega_{n}^{2}},\qquad J_{2}=\sum_{n}J_{p,n}J_{q,n}\frac{1}{\nu^{2}-\Omega_{n}^{2}},\qquad J_{3}=\sum_{n}J_{j,n}J_{q,n}\frac{1}{\nu^{2}-\Omega_{n}^{2}} 
\end{align}
  and that of  the trilinear coupling between three ions
  \begin{align}
  J=\sum_{n}J_{j,n}J_{p,n}h_{n}\frac{1}{\nu^{2}-\Omega_{n}^{2}}.
  \end{align}
    At this level,  it is clearly seen that   one can realize the circulant Hamiltonian  $H_{\sf cir}^{(1)}(t)$
\eqref{4} by adjusting the coupling parameters.
\section{Numerical analysis \label{sec5} }

In what follows, we choose the following  time {modulations} of the couplings $ J_{1}(t) $, $ J(t) $ and the detunings $ \Delta_{j}(t) $ for the gate implementation{:}
\begin{align}
	&J_{1}(t) = J_{01} \sin^{2}(\omega' t)\\
	&  J(t) =J_{0} \sin^{2}(\omega' t)\\
	&
	\Delta_{j}(t) = \Delta_{j}\cos^{2}(\omega' t) 
\end{align}
where  the characteristic parameter $ \omega' $  controls the adiabaticity of the transition and the interaction time $ t $ varies as $ t \in[0,t_{max}] $ with $ t_{max}=\frac{\pi}{2\omega{'}} $. This time dependence guarantees 
{the following} conditions 
\begin{align}
\Delta_{1,2,3}(0)\gg J(0), J_{1}(0), \qquad \Delta_{1,2,3}(t_{max})\ll J(t_{max}), J_{1}(t_{max}).
\end{align}
The adiabatic transition to Fourier modes can be carried {\rr out} without using the detuning $\Delta_{j}$. In fact,  we can simply vary the Rabi frequencies  to finally get  the Fourier modes, such as
\begin{align}
&	J(t)=J_{0}\sin^{2}(\omega' t)\label{Ra1}\\
&	J_{1}(t)=J_{01}\sin^{2}(\omega' t)\label{Ra2}\\
	& \Omega_{2}(t)=J_{01}+\Upsilon_{0}\cos^{2}(\omega' t)\label{Ra3}\\
	& \Omega_{3}(t)=J_{0}+\Upsilon'_{0}\cos^{2}(\omega' t)\label{Ra4}
\end{align}
with $  \Upsilon_{0}$ and $  \Upsilon'_{0}$  are the adding {amplitudes} for the control of the adiabaticity of transition in the second and {third qubits}, respectively.
\subsection{Eigenfrequencies }

\begin{figure}[htbphtbp]
	\centering 
	\includegraphics[width=10cm,height=6cm]{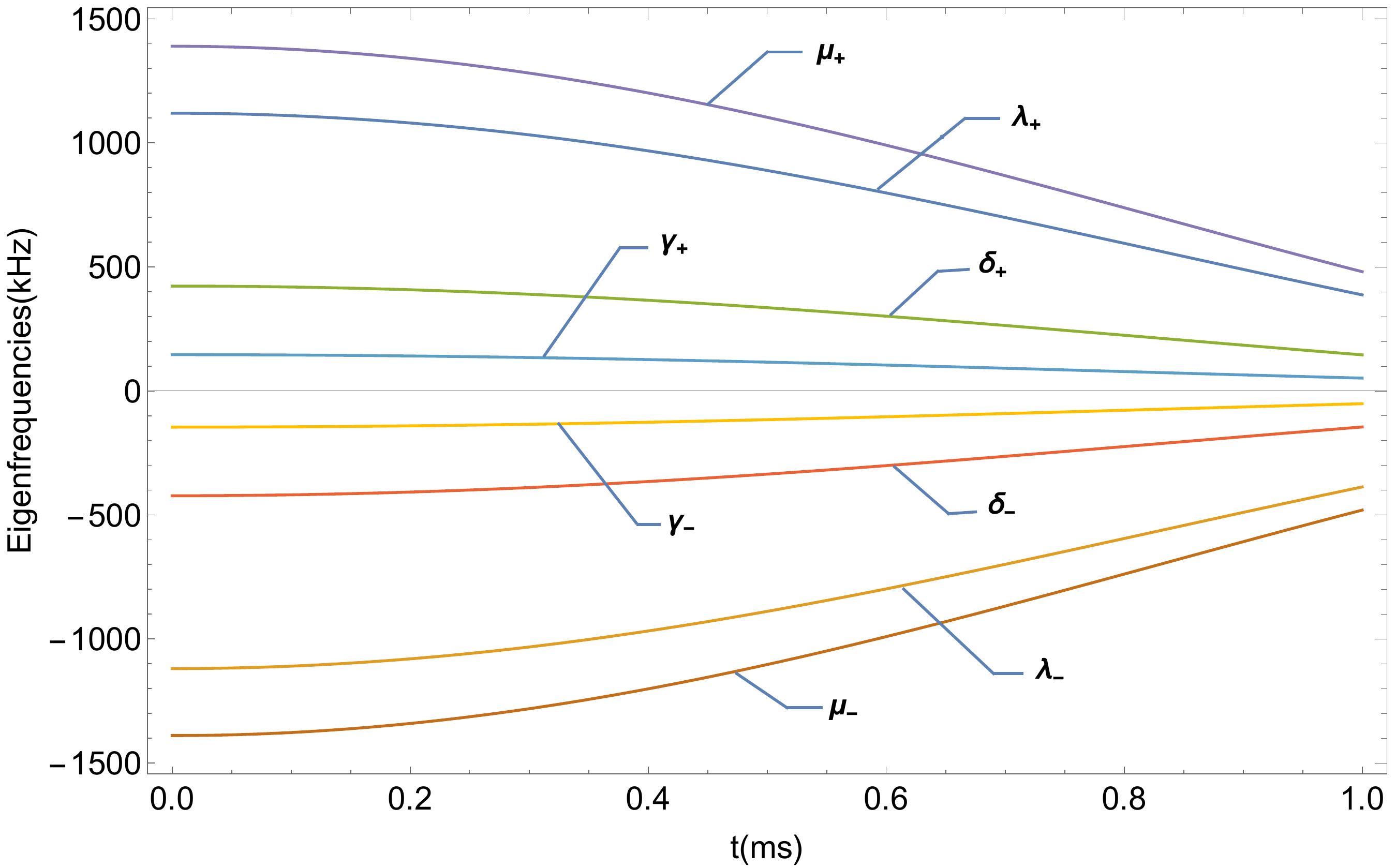} 
	\captionof{figure}{(color online) Eigenfrequencies of 
	the Hamiltonian
$ {H}(t)$ \eqref{Of} 	
	as a function of the time. The parameters are {chosen as}   $ J_{0}/2\pi=1$ kHz, $J_{01}/2\pi= 2$ kHz, $\Delta_{1}/2\pi=120$ kHz, $\Delta_{2}/2\pi= 60$ kHz, $\Delta_{3}/2\pi=30$ kHz, $\varphi=\dfrac{\pi}{2}$, $\omega'/2\pi=0.15$ kHz. }\label{X}
\end{figure}

{We numerically show in {\color{red}Figure} \ref{X} the eigenfrequencies $ \lambda_{\pm}(t),
\delta_{\pm}(t), \mu_{\pm}(t),  \gamma_{\pm}(t) $ of the Hamiltonian $ {H}(t)$ \eqref{Of} 
versus time under suitable choices of the coupling parameters and detunings.} As expected, all eigenfrequencies are separated from each {other}, which entails, in its turn the suppression of any transition to a superposition of eigenstates. The degeneracy of the energies should be avoided during the simulated time, and this is due to the fact that degeneracy entails the prevention of the gate implementation at hand. The gap between the eigenvalues {decreases} during the {simulation} time. {To} avoid their degeneracy during the evolution,  we have introduced the detuning frequencies $ \Delta_j $. Additionally, we mention that the amplitude {couplings} $ J_0 $ and $ J_{01} $ are important to prevent the degeneracy at final time $ t_{max} $. 


\subsection{Gate fidelity }
\begin{figure}[htbphtbp]
	\centering 
    \includegraphics[width=10cm,height=6cm]{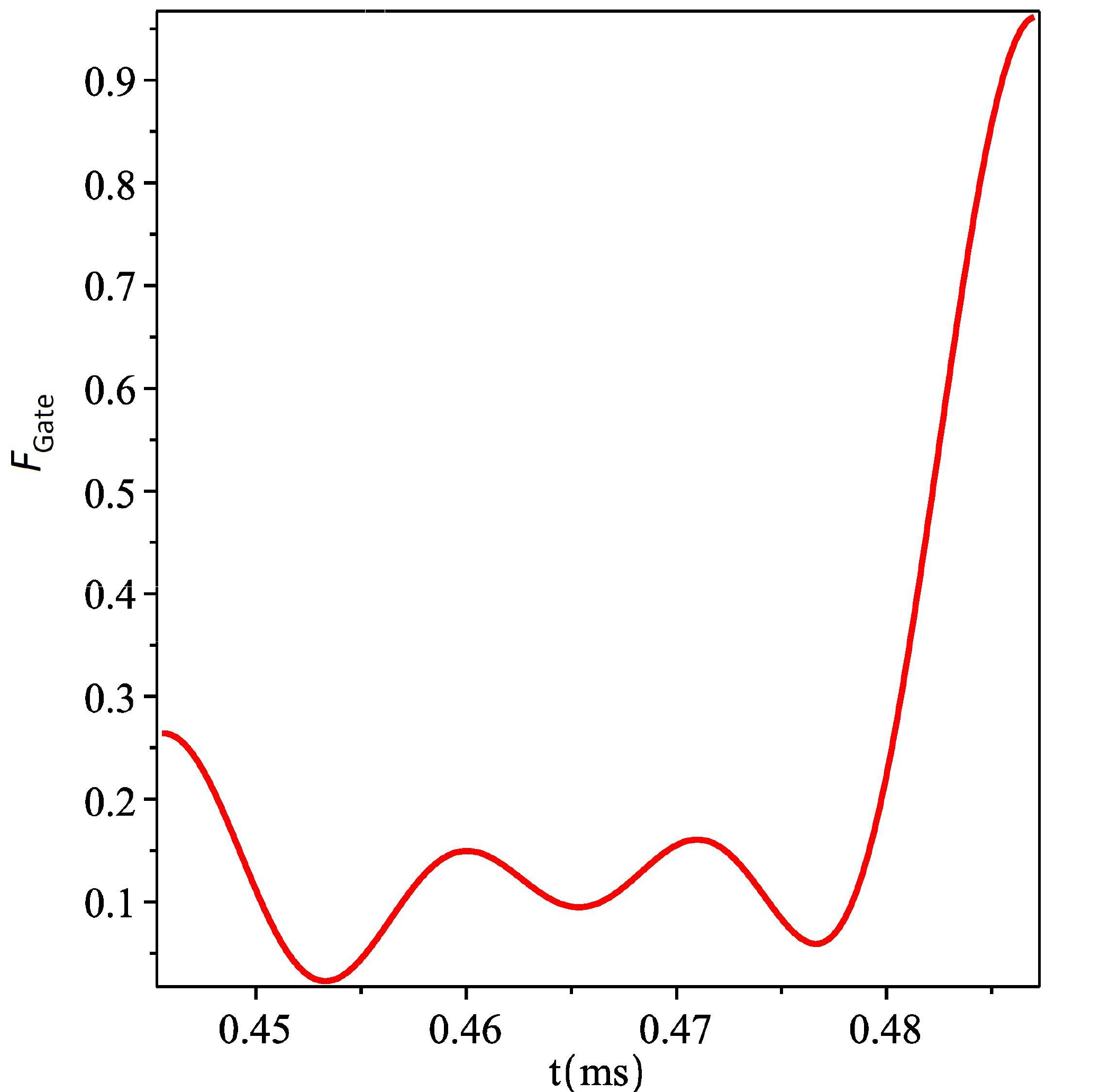} 
	\captionof{figure}{(color online) {The gate fidelity is calculated using the Hamiltonian $ {H}(t)$ \eqref{Of} in a numerical simulation.} 	
	The  parameters are {chosen as}  $ J_{0}/2\pi=J_{01}/2\pi= 1$ kHz, $\Delta_{1}/2\pi=20$ kHz, $\Delta_{2}/2\pi= 10$ kHz, $\Delta_{3}/2\pi=6$ kHz, $\varphi=\dfrac{\pi}{2}$   and $  \omega'=0.505$ kHz. } \label{F}
\end{figure}

{Gate} fidelity is a tool 
to compare how close two gates, or more generally operations, are to each other
\cite{Magesan2011}. In other {words}, 
it expresses the probability that one state will pass a test to identify {itself} as the other one. We recall that 
 fidelities higher than 99.99$\%$ for a single-qubit gate
and 99.9 $\%$ 
for an entangling gate in a two-ion crystal have been
developed in \cite{555,666, 777}. 
Generally, for  {the} theoretical density matrix $\rho_0$ and 
reconstructed density matrix
$\rho$,
it is defined by
\begin{align}\label{frrr}
F(\rho_0, \rho)= \left( \Tr \sqrt{\sqrt\rho_0 \rho\sqrt\rho_0}\right)^2.
\end{align}
By applying the Uhlmann theorem \cite{3737}, \eqref{frrr} can take a simple form
\begin{align}\label{frrr2}
F(\rho_0, \rho)= \left|\langle\psi_0|\psi\rangle \right|^2.
\end{align}
with $\psi_0$ and $\psi$ are theoretical and reconstructed purified state vectors, {respectively}.
As for our system, we have 
  \cite{circ2} 
\begin{eqnarray}
	F_{\sf Gate}(t) = \dfrac{1}{16}\left|\sum_{s_{1},s_{2},s_{3}}\bra{s_{1}s_{2}s_{3}}\mathcal{G}^{+}\mathcal{G}'(t)\ket{s_{1}s_{2}s_{3}}\right|^{2}
\end{eqnarray}
where $ s_{j} = \uparrow_{j},\downarrow_{j} $, $ \mathcal{G} $ is the three-qubit QFT (\ref{G}) and $ \mathcal{G}'(t) $ is the real transform.  
In {\color{red}Figure} \ref{F}, we present the gate fidelity versus the evolution time  by choosing the detunings $ \Delta_{1,2,3} $ such that the adiabatic phases are given in
\eqref{phases}.
The unitary propagator $ \mathcal{G}'(t) $ converges to $ \mathcal{G} $ as time progresses. We notice that for a nonlinear coupling $ J_{0} = 1$ kHz  and $ t = 0.4875$ ms,  the gate reaches a high fidelity ($96$\textdiscount). 

By using the Hamiltonian  (\ref{588})
 together with  the quantum Fourier states (\ref{3}-\ref{11}), one can end up with
the fidelity of the adiabatic transitions between the rotating computational spin states $ \ket{s'_{1}s'_{2}s'_{3}} (s'_{j}=\pm_{j}) $ 
\begin{eqnarray}
F_{\sf ad}(t) = \dfrac{1}{16}\left|\sum_{i=0}^{i=7} \bra{\psi_{i}}\ket{\Lambda_{i}} \right|^{2}
\end{eqnarray}
and 
more explicitly, we have
\begin{align}
F_{\sf ad}(t) =& \dfrac{1}{16}\left|\bra{ \psi_{0}}\ket{\Lambda_{0}} +\bra{\psi_{1}}\ket{\Lambda_{1}}+\bra{\psi_{2}}\ket{\Lambda_{2}}+\bra{\psi_{3}}\ket{\Lambda_{3}}+\bra{\psi_{4}}\ket{\Lambda_{4}}+\bra{\psi_{5}}\ket{\Lambda_{5}}+\bra{\psi_{6}}\ket{\Lambda_{6}}+\bra{\psi_{7}}\ket{\Lambda_{7}}\right|^{2}.
\end{align}
\begin{figure}[htbphtbp]
\centering 
\includegraphics[width=10cm,height=6cm]{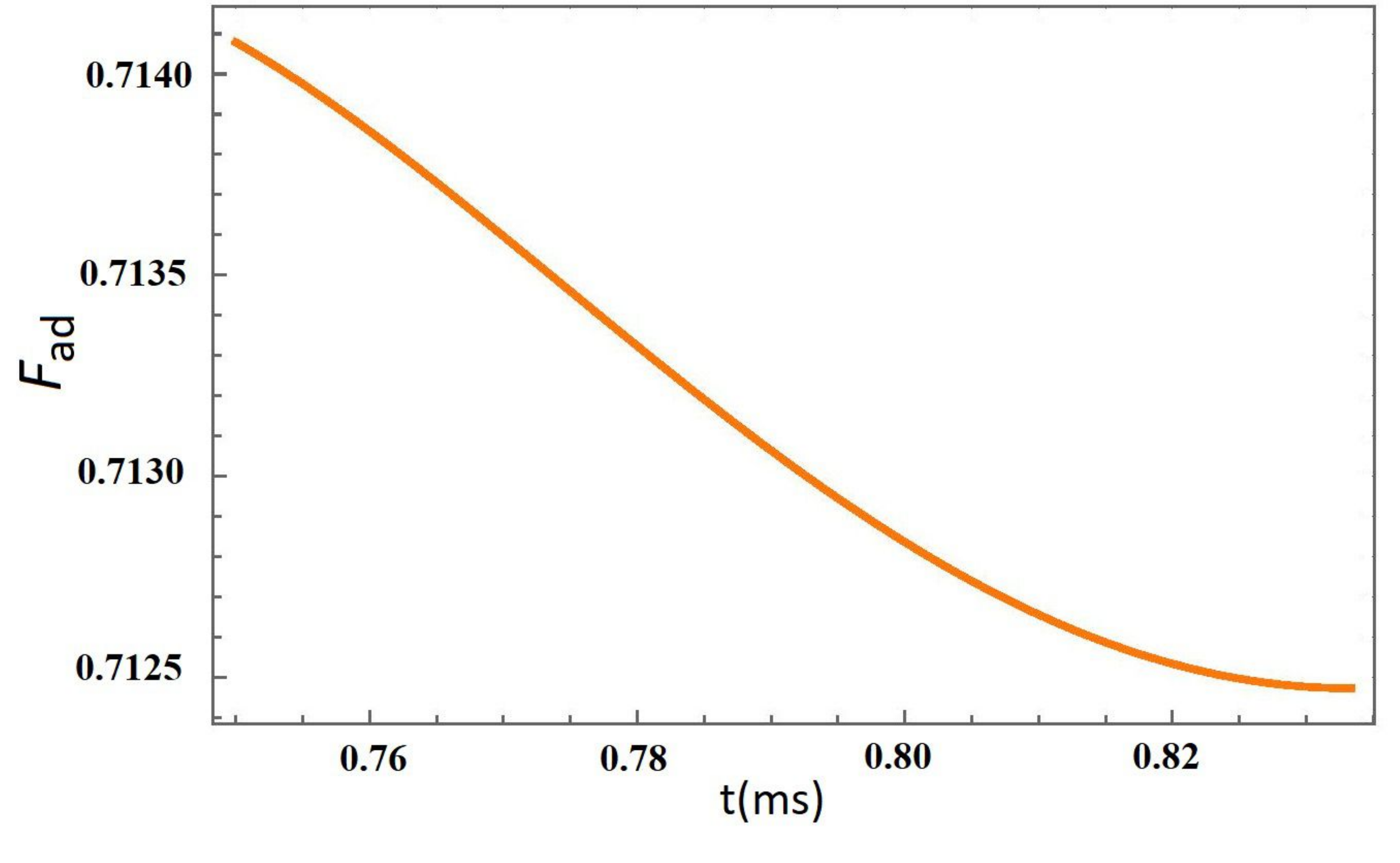} 
\captionof{figure}{(color online) Fidelity of adiabatic transition with $ J_{0}/2\pi = 2.1$ kHz, $ \Upsilon_{0}/2\pi = 1.9$ kHz, $ J_{01}/2\pi = 2.4$ kHz, $ \Upsilon'_{0}/2\pi = 2$ kHz, $ \varphi=\dfrac{\pi}{4} $ and $ \omega'/2\pi = 0.3$ kHz. } \label{A}
\end{figure}

{\color{red}Figure} \ref{A} presents the good fidelity of the adiabatic transition within a shorter interaction time, $ t_{max} = 0.835$ ms. It is clearly seen that  our  results show the possibility {for obtaining} high fidelity ($71$\textdiscount).

\subsection{Creation of entangled states}

\begin{figure}[H]
	\centering 
	\includegraphics[width=5.5cm,height=4cm]{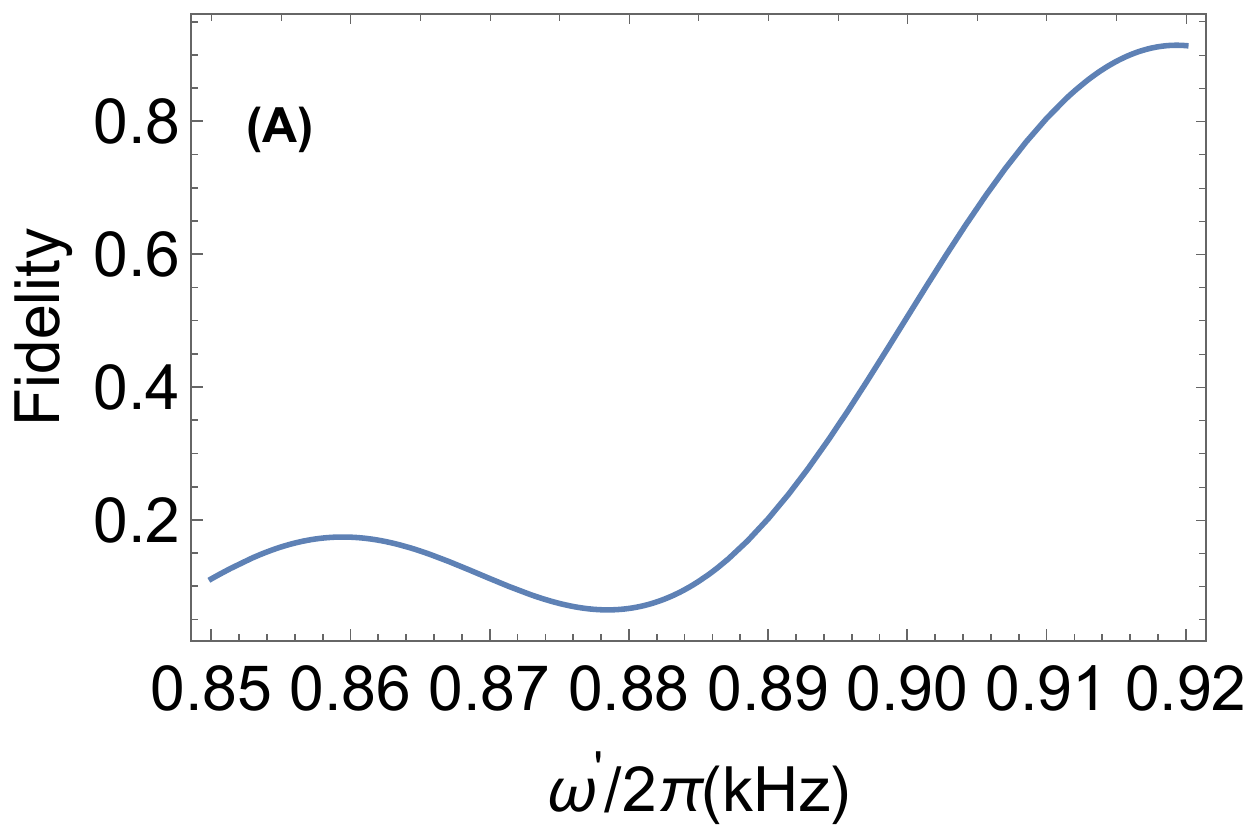}
	\includegraphics[width=5.5cm,height=4cm]{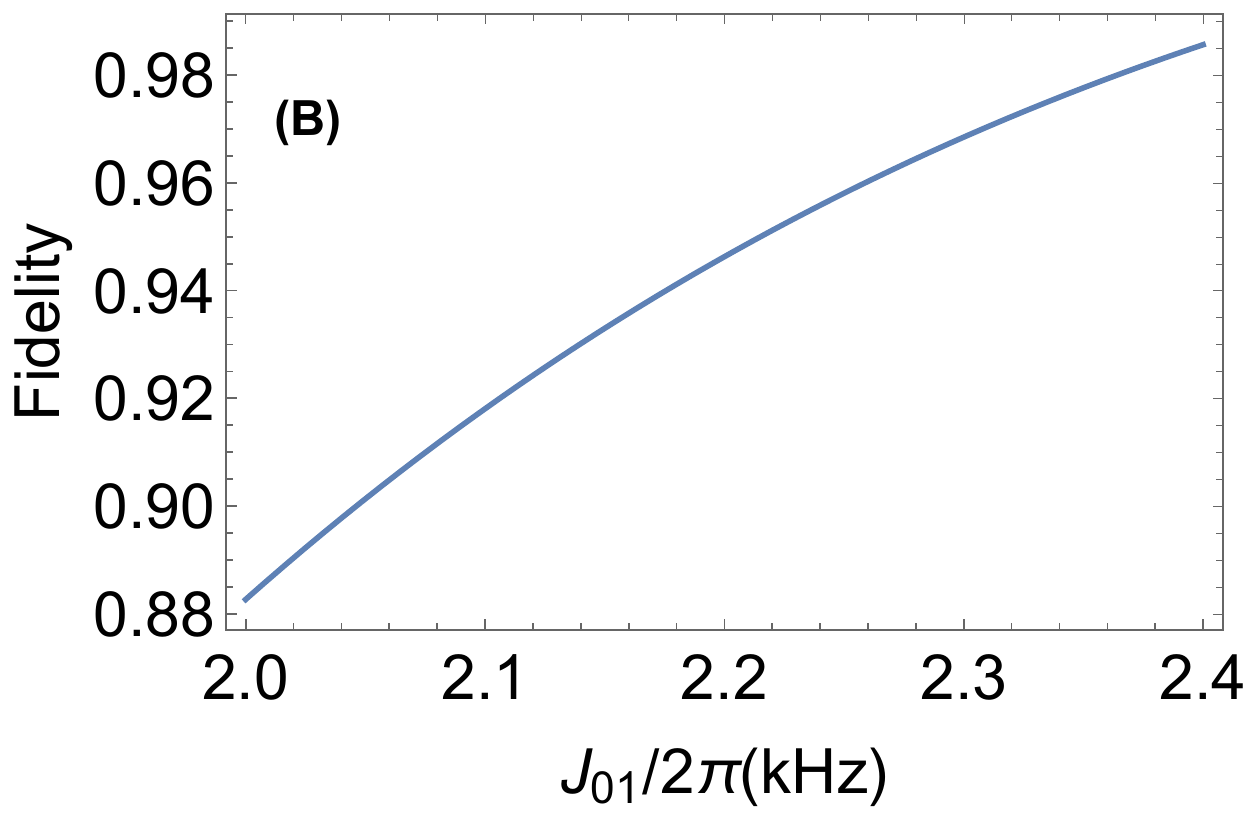}
	\includegraphics[width=5.5cm,height=4cm]{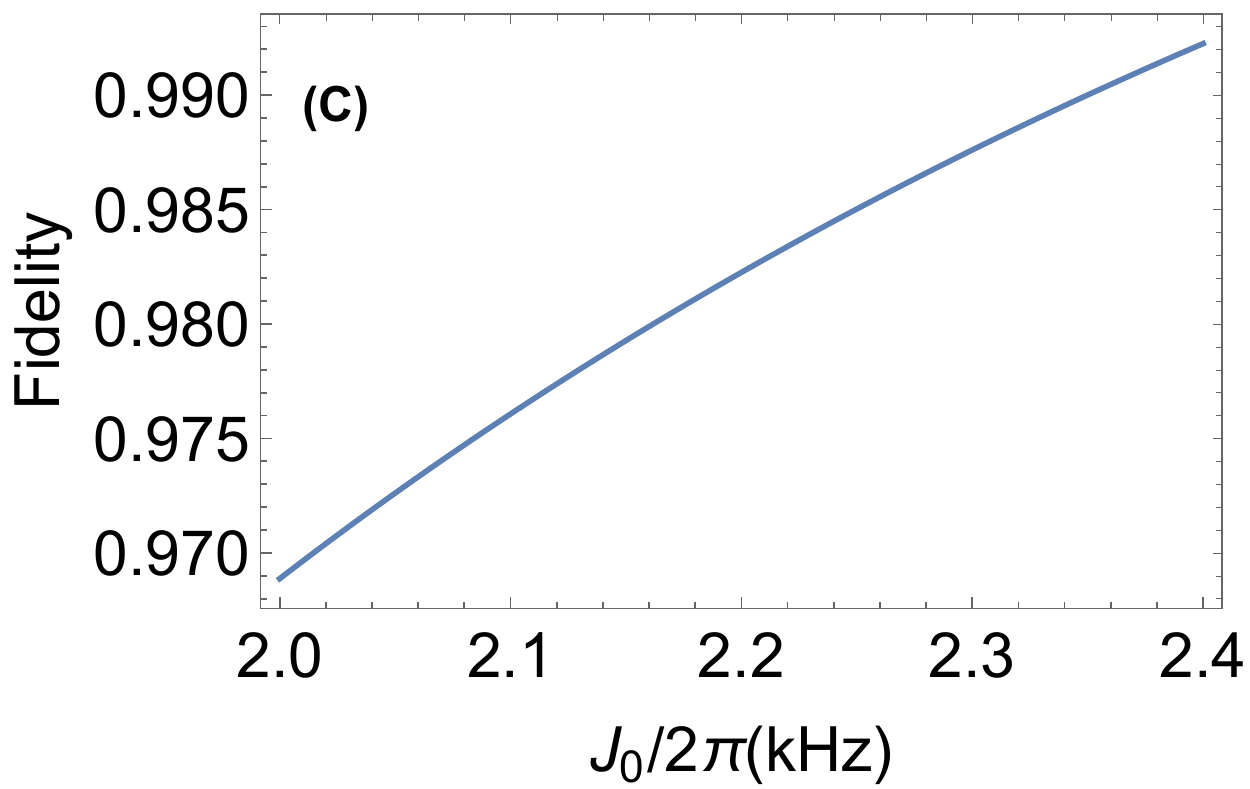}  
	\captionof{figure}{(color online) {The fidelity of the entangled state is calculated}  from the numerical simulation of the Hamiltonian \eqref{588}. (A): $ \Upsilon_{0} = 1.8$ kHz,$\Upsilon'_{0} = 1.7$ kHz, $J_{01} = 2.1$ kHz, $J_{0} = 2.3$ kHz  and the gate time $ t= 0.31$ ms. (B): The same values with $ \omega'/2\pi=0.5$ kHz and vary the coupling strength $ J_{01} $. (C): The same values in (A) with $ \omega'/2\pi=0.605$ kHz  and vary the coupling strength $ J_{0} $.  }
\label{B}
\end{figure}

To create entangled states, one has to suitably prepare the initial state in a superposition of spin states, which is due to the fact that the action of the QFT on the computational basis creates a superposition, but they are not entangled. For concreteness, we assume that the system is initially prepared in  the following state{\rr :}
\begin{eqnarray}
	\ket{\psi(0)}= \frac{1}{2} \ket{\downarrow_{1}}(e^{i\alpha_{1}}\ket{\downarrow_{2}\downarrow_{3}} +e^{i\alpha_{2}}\ket{\downarrow_{2}\uparrow_{3}}+e^{i\alpha_{3}}\ket{\uparrow_{2}\downarrow_{3}}+e^{i\alpha_{4}}\ket{\uparrow_{2}\uparrow_{3}}). 
\end{eqnarray}
Performing our three {qubit} gates, we obtain the  entangled state
\begin{eqnarray}
	\ket{\psi(0)} \longrightarrow \ket{\psi(t_{f})} = \frac{1}{2}(\ket{\psi_{0}}+\ket{\psi_{1}}+\ket{\psi_{2}}+\ket{\psi_{3}}).\label{entang}
\end{eqnarray}
Let us {emphasize} that  by rotating the initially prepared state such as
\begin{eqnarray}
	\ket{\psi(0)}= \frac{1}{2} \ket{-_{1}}(e^{-i\beta_{0}}\ket{-_{2}-_{3}} +e^{-i\beta_{1}}\ket{-_{2}+_{3}}+e^{-i\beta_{2}}\ket{+_{2}-_{3}}+e^{-i\beta_{3}}\ket{+_{2}+_{3}})
\end{eqnarray}
we end up with   the same transformed entangled state (\ref{entang}).  {Therefore}, the fidelity of the creation of the entangled state is defined by 
\begin{eqnarray}
	F(t) = \dfrac{1}{2}|\bra{\psi(t_{f})}(e^{-i\beta_{0}}\ket{\Lambda_{0}(t)}+e^{-i\beta_{1}}\ket{\Lambda_{1}(t)}+e^{-i\beta_{2}}\ket{\Lambda_{2}(t)}+e^{-i\beta_{3}}\ket{\Lambda_{3}(t)})|^{2}.	
\end{eqnarray}
 By adjusting the parameters  $\omega'$, $J_{01}$ and $J_{0}$ one can {achieve} high fidelity {\rr in} the creation of entangled states as presented in  {\color{red}Figure \ref{B}-A}, {\color{red}Figure \ref{B}-B} and  {\color{red}Figure \ref{B}-C}.
 
\section{Short-cut to adiabaticity \label{sec6} }
Now we add an  auxiliary interaction $ {H}_{\sf CD}(t) $ (counter-driving-field) \cite{HCD,Adia} to the reference Hamiltonian $ {H}^{(1)}(t) $ \eqref{588} in order to suppress the non-adiabatic transitions and reduce the gate time. As a result, the Hamiltonian will take {the following form:}
\begin{equation}
{H}_{\sf T}(t)={H}^{(1														)}(t)+{H}_{\sf CD}(t) \label{66}
\end{equation}
such that the interaction is 
\begin{eqnarray}
	{H}_{\sf CD}(t) = i\hbar \sum_{i=0}^{7}\ket{\partial_{t}\Lambda_{i}(t)}\bra{\Lambda_{i}(t)} \label{67}
\end{eqnarray}
and  {(\ref{64}-\ref{71}) show  the time-dependent eigenvectors $ \ket{\Lambda_{i}(t)} $ 
of ${H}^{(1)}(t) $}. After some algebra, we obtain
\begin{eqnarray}
	{H}_{\sf CD}(t)=-
	\partial_{t}\kappa(t)\begin{pmatrix}
		0&0&0&0&1&0&0&0 \\
		0& 0 
		&0&0&0& 1 
		&0&0  \\
		0&0&0&0&0&0&0&0  \\
		0&0&0&0&0&0&0&0 \\
		1&0&0&0&0&0&0&0 \\
		0& 1 
		&0&0&0& 0 
		&0&0 \\
		0&0&0&0&0&0&0&0 \\
		0&0&0&0&0&0&0&0
	\end{pmatrix}.
\end{eqnarray} 
Using the time-dependent coupling parameters  (\ref{Ra1}-\ref{Ra4}) and the eigenvectors (\ref{64}-\ref{71}), we explicitly determine the counter-driving-field
\begin{align}\label{72}
{\partial_{t}\kappa(t) = \dfrac{1}{2}\left(
 \dfrac{\omega' J_{01}(J_{01}+\Upsilon_{0})\sin(2\omega't)}{J_{01}^{2}\sin^{4}(\omega' t)+[\Upsilon_{0}\sin^{2}(\omega' t)-(J_{01}+\Upsilon_{0})]^{2}} + \dfrac{\omega' J_{0}(J_{0}+\Upsilon_{0}')\sin(2\omega't)}{J_{0}^{2}\sin^{4}(\omega' t)+[\Upsilon_{0}'\sin^{2}(\omega' t)-(J_{0}+\Upsilon_{0}')]^{2}}\right). }
\end{align}

\begin{figure}[H]
	\centering 
	\includegraphics[width=9cm,height=4.8cm]{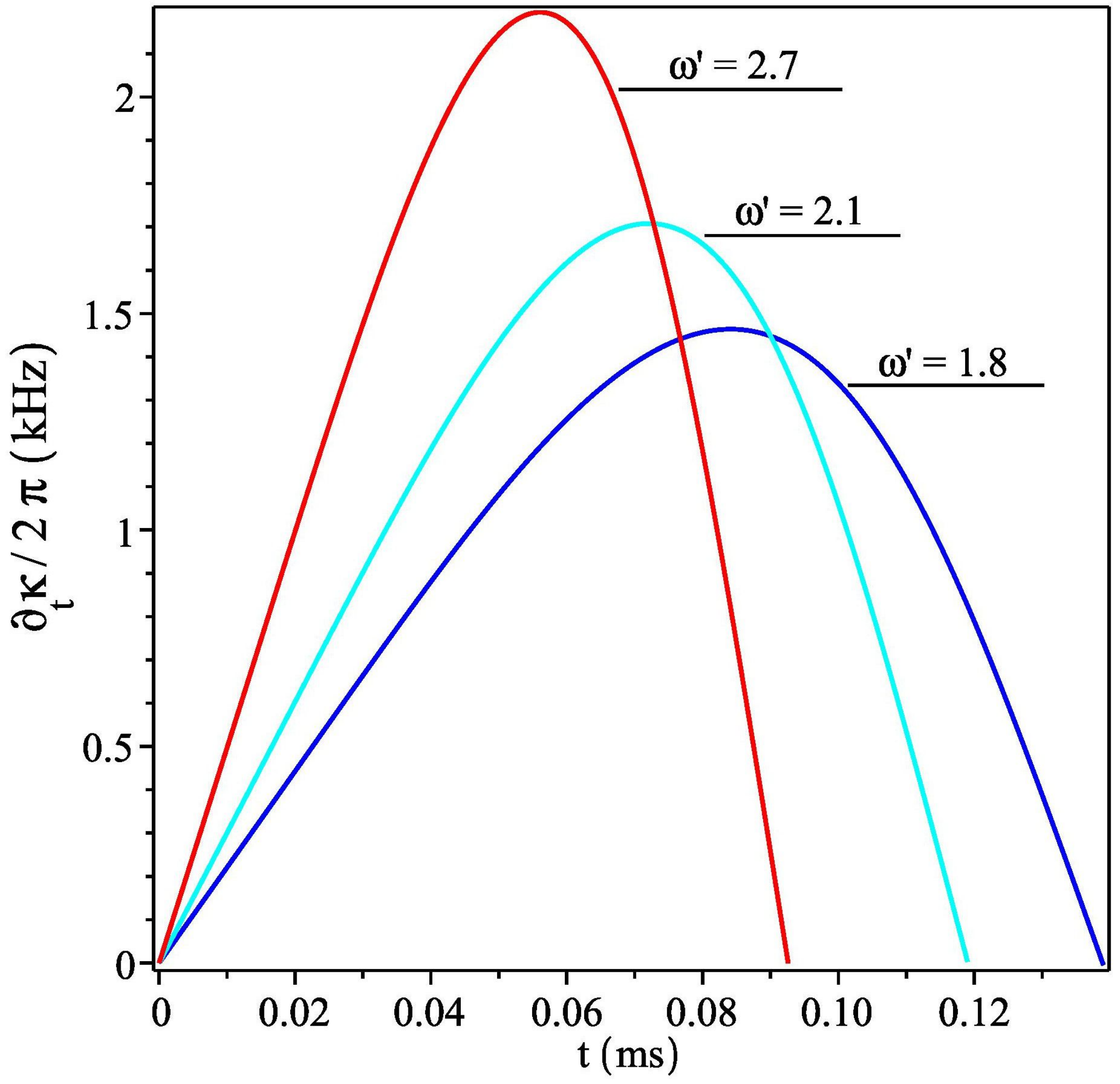} 
	\captionof{figure}{(color online) Behavior  of {a} counter-driving field  with $\Upsilon_{0}/2\pi= 0.5$ kHz  and $\Upsilon'_{0}/2\pi= 2$ kHz, {while the remaining} parameters are {chosen as} $ J_{01}=1.5$ kHz, $J_{0}=1$ kHz  {(blue line)}, $ J_{01}=1.9$ kHz, $J_{0}=1.3$ kHz  {(cyan line)},$ J_{01}=2$ kHz, $J_{0}=1.7$ kHz {(red line)}.}  \label{Y}
\end{figure}

In {\color{red} Figure \ref{Y}}, we show the shape of the counter-driving field (\ref{72}) as a function of the time and by varying $ \omega' $,  $ J_{01} $ and $ J_{0} $. The counter-driving term should be zero at $ t=0 $, because the system starts in the {rotational} computational spin states. We mention also that at $ t_{max} $, the system {ends} up with the Fourier modes.

\section{Conclusion}

We generalized the work done by  Ivanov and Vitanov \cite{circ2}
dealing with two-qubit quantum gates and
entanglement protected by
circulant symmetry to a system of three-qubit quantum gates.
In fact, 
we have constructed a discrete system based on three qubits {emerging} in a magnetic field. A special  symmetry called circulant {can be} obtained only by adjusting the Rabi frequencies and the coupling parameters characterizing our system. {We have shown that our  eigenvectors do not depend on the magnitude of the physical parameters}, which entails the protection of entanglement. {These  eigenvectors lead to  obtaining the 
quantum Fourier transform
(QFT) modes}, which imply the realization of a QFT gate. To discuss the implementation of the gate,   the eigenfrequencies should be non-degenerate. {For this aim, we have  added an Hamiltonian $ H_{0}(t) $} that breaks the circulant symmetry and favors the adiabatic transition process.  

Subsequently,  instead of adding an energy offset, we have  shown that it is possible to  control the transition by using only the  Rabi frequencies.  
{As a result, the gate scheme 
with the short-cut to adiabaticity has been discussed, and therefore} we have found  the  suppression of the non-adiabatic transition and {the acceleration of}  the gate. {Additionally,  in the framework of the trapped ions}, we have suggested a
possible   physical implementation of the constructed Hamiltonian.
 By assuming a particular sinusoidal modulation, several fidelities are discussed, and the results show the possibility to achieve high fidelities only by adjusting the physical parameters. 
 The physical realization of a three-qubit QFT is the key subroutine in several quantum algorithms. {Then,} it turns out that our  present three-qubit gates can significantly  reduce the number of {gates} in a quantum algorithm.
 
 {In this work, we have added the trilinear coupling, which is necessary to retain the circulant symmetry in a three-qubit system. Indeed, a three-qubit system with only bilinear interaction terms, as done in \cite{circ1}, does not lead to the wanted circulant symmetry. Additionally, constructing a circulant Hamiltonian based on $ N $-qubits is not obvious. Indeed, the Hamiltonian demands multilinear interaction terms that are not easily realizable. Moreover, discussing the adiabatic transition and accelerating the gate using a shortcut to adiabacity will be hard.}

\newpage
\begin{appendices}
\numberwithin{equation}{section}
\section{Eigenfrequencies of $ {H}(t) $ with energy offset}
\label{appendixA}
{As for $ {H}(t) $ \eqref{Of}   including the circulant Hamiltonian \eqref{4} and the energy offset
\eqref{hdet}, we show that the corresponding eigenfrequencies are given by}
\begin{align} 
	&\lambda_{\pm}=\pm \sqrt{\left|-\frac{A}{4}-S+ \frac{1}{2}\sqrt{\left|-4S^{2}-2p+\frac{q}{S}\right|}\right|}\label{a1}\\
	& \delta_{\pm}=\pm\sqrt{\left|-\frac{A}{4}-S- \frac{1}{2}\sqrt{\left|-4S^{2}-2p+\frac{q}{S}\right|}\right|}\label{a2}\\
&	\mu_{\pm}=\pm\sqrt{\left|-\frac{A}{4}+S+ \frac{1}{2}\sqrt{\left|-4S^{2}-2p-\frac{q}{S}\right|}\right|}\label{a3}\\
&
	\gamma_{\pm}=\pm\sqrt{\left|-\frac{A}{4}+S-\frac{1}{2}\sqrt{\left|-4S^{2}-2p-\frac{q}{S}\right|}\right|}\label{a4}
\end{align} 
where we have set
\begin{align}
	& p = \frac{1}{8} \left(8B-3\right)\\
	&  q = \frac{1}{8} \left(-1+4B+8C\right) \\
	&  S = \frac{1}{2} \sqrt{\frac{1}{3}\left|-2p +\left(Q+\frac{\Delta_{0}}{Q}\right)\right|}\\
	& \Delta_{0} = B^{2} + 3C + 12D\\
	& Q = \sqrt[3]{\frac{1}{2}\left|\Delta+\sqrt{\left|\Delta^{2}-4\Delta_{0}^{3}\right|}\right|}\\
	&   \Delta = 2B^{3} + 9BC + 27D + 27C^{2} - 72BD 	
\end{align}
and the {associated} quantities are
\begin{align}\nonumber
		A=&-16J^{2} - 4\Delta_{1}^{2} - 4\Delta_{2}^{2} - 4\Delta_{3}^{2} - 8J_{1}^{2}\\ \nonumber
		B=&32 {J}^{2}{\Delta_{{1}}}^{2}+32\,{J}^{2}{\Delta_{{2}}}^{2}+48 {J}^{2}{\Delta_{{3}}}^{2}+128 {J}^{2}{J_{{1}}}^{2}+6{\Delta_{{1}}}^{4}+4\,{\Delta_{{1}}}^{2}{\Delta_{{2}}}^{2}+4{\Delta_{{1}}}^{2}{\Delta_{{3}}}^{2}\nonumber\\
	&	+16{\Delta_{{1}}}^{2}{J_{{1}}}^{2}+6\,{\Delta_{{2}}}^{4}
	+4{\Delta_{{2}}}^{2}{\Delta_{{3}}}^{2}+24{\Delta_{{2}}}^{2}{J_{{1}}}^{2}+6{\Delta_{{3}}}^{4}+8{\Delta_{{3}}}^{2}{J_{{1}}}^{2}+16{J_{{1}}}^{4}\nonumber\\ 	\nonumber
		C=&-16{J}^{2}{\Delta_{{1}}}^{4}-32{J}^{2}{\Delta_{{1}}}^{2}{\Delta_{{2}}}^{2}-128{J}^{2}{\Delta_{{1}}}^{2}{J_{{1}}}^{2}-16{J}^{2}{\Delta_{{2}}}^{4}-128{J}^{2}{\Delta_{{2}}}^{2}{J_{{1}}}^{2}\nonumber\\
&		-48{J}^{2}{\Delta_{{3}}}^{4}-256\,{J}^{2}{J_{{1}}}^{4} -4{\Delta_{{1}}}^{6}+4{\Delta_{{1}}}^{4}{\Delta_{{2}}}^{2}+4{\Delta_{{1}}}^{4}{\Delta_{{3}}}^{2}-8{\Delta_{{1}}}^{4}{J_{{1}}}^{2}+4{\Delta_{{1}}}^{2}{\Delta_{{2}}}^{4}\nonumber\\
&
-40{\Delta_{{1}}}^{2}{\Delta_{{2}}}^{2}{\Delta_{{3}}}^{2}+4{\Delta_{{1}}}^{2}{\Delta_{{3}}}^{4}-32{\Delta_{{1}}}^{2}{\Delta_{{3}}}^{2}{J_{{1}}}^{2}-4\,{\Delta_{{2}}}^{6}+4{\Delta_{{2}}}^{4}{\Delta_{{3}}}^{2}-24\,{\Delta_{{2}}}^{4}{J_{{1}}}^{2}\nonumber\\
&+4{\Delta_{{2}}}^{2}{\Delta_{{3}}}^{4}+16\,{\Delta_{{2}}}^{2}{\Delta_{{3}}}^{2}{J_{{1}}}^{2}-32\,{\Delta_{{2}}}^{2}{J_{{1}}}^{4}-4\,{\Delta_{{3}}}^{6}+8\,{\Delta_{{3}}}^{4}{J_{{1}}}^{2}-32\,{\Delta_{{3}}}^{2}{J_{{1}}}^{4}
		\nonumber\\
		D=&4\,{\Delta_{{1}}}^{4}{\Delta_
			{{2}}}^{2}{\Delta_{{3}}}^{2}+4\,{\Delta_{{1}}}^{2}{\Delta_{{2}}}^{4}{\Delta_{{3}}}^{2}+4\,{\Delta_{{1}}}^{2}{\Delta_{{2}}}^{2}{\Delta_{{3}}}^{4}-4\,{\Delta_{{1}}}^{6}{\Delta_{{2}}}^{2}-4\,{\Delta_{{1}}}^{6}{\Delta_{{3}}}^{2}+6\,{\Delta_{{1}}}^{4}{\Delta_{{2}}}^{4}\nonumber\\ \nonumber
		&+6\,{\Delta_{{1}}}^{4}{\Delta_{{3}}}^{4}-4\,{\Delta_{{1}}}^{2}{\Delta_{{2}}}^{6}-4\,{\Delta_{{1}}}^{2}{\Delta_{{3}}}^{6}-4\,{\Delta_{{2}}}^{6}{\Delta_{{3}}}^{2}+6\,{\Delta_{{2}}}^{4}{\Delta_{{3}}}^{4}-4\,{\Delta_{{2}}}^{2}{\Delta_{{3}}}^{6}+16\,{\Delta_{{2}}}^{4}{J_{{1}}}^{4}\nonumber\\
		&-8{\Delta_{{3}}}^{6}{J_{{1}}}^{2} 
		+16\,{\Delta_{{3}}}^{4}{J_{{1}}}^{4}+16\,{J}^{2}{\Delta_{{3}}}^{6}+8\,{\Delta_{{2}}}^{6}{J_{{1}}}^{2}+16\,{\Delta_{{1}}}^{2}{\Delta_{{3}}}^{4}{J_{{1}}}^{2}-24\,{\Delta_{{2}}}^{4}{\Delta_{{3}}}^{2}{J_{{1}}}^{2}\nonumber\\
		&+24\,{\Delta_{{2}}}^{2}{\Delta_{{3}}}^{4}{J_{{1}}}^{2} 
		-32\,{\Delta_{{2}}}^{2}{\Delta_{{3}}}^{2}{J_{{1}}}^{4} +
		8{\Delta_{{1}}}^{4}{\Delta_{{2}}}^{2}{J_{{1}}}^{2}-8\,{\Delta_{{1}}}^{4}{\Delta_{{3}}}^{2}{J_{{1}}}^{2}-16\,{\Delta_{{1}}}^{2}{\Delta_{{2}}}^{4}{J_{{1}}}^{2} \nonumber
		\\
		&-128\,{J}^{2}{\Delta_{{3}}}^{4}{J_{{1}}}^{2}	+256\,{J}^{2}{\Delta_{{3}}}^{2}{J_{{1}}}^{4}+16\,{J}^{2}{\Delta_{{1}}}^{4}{\Delta_{{3}}}^{2} -32\,{J}^{2}{\Delta_{{1}}}^{2}{\Delta_{{3}}}^{4}+16\,{J}^{2}{\Delta_{{2}}}^{4}{\Delta_{{3}}}^{2}\nonumber\\
		&-32\,{J}^{2}{\Delta_{{2}}}^{2}{\Delta_{{3}}}^{4}+32\,{J}^{2}{\Delta_{{1}}}^{2}{\Delta_{{2}}}^{2}{\Delta_{{3}}}^{2}+128\,{J}^{2}{\Delta_{{1}}}^{2}{\Delta_{{3}}}^{2}{J_{{1}}}^{2}+128\,{J}^{2}{\Delta_{{2}}}^{2}{\Delta_{{3}}}^{2}{J_{{1}}}^{2} 
		+{\Delta_{{1}}}^{8}+{\Delta_{{2}}}^{8}+{\Delta_{{3}}}^{8} \nonumber.
	\end{align} 
	
	\newpage
\section{Energy spectrum of $ H^{(1)}(t) $}
\label{appendixB}
The Hamiltonian $ H^{(1)}(t) $ with $ \Omega_{2}(t)  $ (i.e. $ J_{1}=\Omega_{2}(t) $ and $ J=\Omega_{3}(t) $ are not always respected) {has the following form:} 
\begin{align} \label{ham1}
	\nonumber
	{H}^{(1)}(t)=&J_{1}(\sigma_{1}^{+} +\sigma_{1}^{-})(\sigma_{2}^{+} e^{-i\varphi}+\sigma_{2}^{-} e^{i\varphi})+\Omega_{3}(t)(\sigma_{2}^{+} +\sigma_{2}^{-})(\sigma_{3}^{+} e^{-i\varphi}+\sigma_{3}^{-} e^{i\varphi})\\ 
	&+\Omega_{3}(t)(\sigma_{1}^{+} +\sigma_{1}^{-})(\sigma_{3}^{+} e^{i\varphi}+\sigma_{3}^{-} e^{-i\varphi})+\Omega_{2}(t)(\sigma_{2}^{+}e^{i\varphi}+\sigma_{2}^{-}e^{-i\varphi})\\  
	&+\Omega_{3}(t)(\sigma_{3}^{+}e^{i\varphi}+\sigma_{3}^{-}e^{-i\varphi})
	+J(\sigma_{1}^{+} +\sigma_{1}^{-} )(\sigma_{2}^{+} +\sigma_{2}^{-} )(\sigma_{3}^{+} e^{-i\varphi}+\sigma_{3}^{-} e^{i\varphi})
	\nonumber
\end{align}
{and we have in the matrix notation}
\begin{align}\label{588}
&{H}^{(1)}(t)=\\
&
\begin{pmatrix}
0&\Omega_{3}(t)e^{i\varphi}&\Omega_{2}(t)e^{i\varphi}&\Omega_{3}(t)e^{-i\varphi}&0&\Omega_{3}(t)e^{i\varphi}&J_{1}e^{-i\varphi}&Je^{-i\varphi}  \\
\Omega_{3}(t)e^{-i\varphi}&0&\Omega_{3}(t)e^{i\varphi}&\Omega_{2}(t)e^{i\varphi}&\Omega_{3}(t)e^{-i\varphi}&0&Je^{i\varphi}&J_{1}e^{-i\varphi}\\ 
\Omega_{2}(t)e^{-i\varphi}&\Omega_{3}(t)e^{-i\varphi}&0&\Omega_{3}(t)e^{i\varphi}&J_{1}e^{i\varphi}&Je^{-i\varphi}&0&\Omega_{3}(t)e^{i\varphi} \\
\Omega_{3}(t)e^{i\varphi}&\Omega_{2}(t)e^{-i\varphi} & \Omega_{3}(t)e^{-i\varphi} & 0 & Je^{i\varphi} & J_{1}e^{i\varphi} & \Omega_{3}(t)e^{-i\varphi} & 0 \\
0	&\Omega_{3}(t)e^{i\varphi} & J_{1}e^{-i\varphi} & Je^{-i\varphi}& 0 & \Omega_{3}(t)e^{i\varphi} &  \Omega_{2}(t)e^{i\varphi}& \Omega_{3}(t)e^{-i\varphi} \\
\Omega_{3}(t)e^{-i\varphi}	&0& Je^{i\varphi} & J_{1}e^{-i\varphi} & Je^{-i\varphi}& 0 & \Omega_{3}(t)e^{i\varphi} &  \Omega_{2}(t)e^{i\varphi}\\
J_{1}e^{i\varphi}&Je^{-i\varphi}&0&\Omega_{3}(t)e^{i\varphi}&\Omega_{2}(t)e^{-i\varphi}&\Omega_{3}(t)e^{-i\varphi}&0&\Omega_{3}(t)e^{i\varphi}  \\
Je^{i\varphi}&J_{1}e^{i\varphi}&\Omega_{3}(t)e^{-i\varphi}&0&\Omega_{3}(t)e^{i\varphi}&\Omega_{2}(t)e^{-i\varphi}&\Omega_{3}(t)e^{-i\varphi} & 0
			\end{pmatrix} \nonumber.
\end{align} 
{We choose $ \varphi = \dfrac{\pi}{4} $ 
for simplicity 
and show that  the  time-dependent eigenfrequencies 
$ {H}^{(1)}(t) $ \eqref{588} are given by}
\begin{align}
&	\Lambda_{0}(t)=\sqrt{(\Omega_{3}-J)^{2}+J_{1}^{2}+\Omega_{2}^{2}+\sqrt{2(\Omega_{3}-J)^{2}(J_{1}-\Omega_{2})^{2}}} \label{bb8}\\
&	\Lambda_{1}(t)=-\sqrt{(\Omega_{3}-J)^{2}+J_{1}^{2}+\Omega_{2}^{2}+\sqrt{2(\Omega_{3}-J)^{2}(J_{1}-\Omega_{2})^{2}}}\label{bb7}\\
&	\Lambda_{2}(t)=\sqrt{(\Omega_{3}-J)^{2}+J_{1}^{2}+\Omega_{2}^{2}-\sqrt{2(\Omega_{3}-J)^{2}(J_{1}-\Omega_{2})^{2}}}\label{bb6}\\
&	\Lambda_{3}(t)=-\sqrt{(\Omega_{3}-J)^{2}+J_{1}^{2}+\Omega_{2}^{2}-\sqrt{2(\Omega_{3}-J)^{2}(J_{1}-\Omega_{2})^{2}}}\label{bb5}\\
&	\Lambda_{4}(t)= \sqrt{5\Omega_{3}^{2}+2J\Omega_{3}+J^{2}+J_{1}^{2}+\Omega_{2}^{2}+\sqrt{2\Omega_{3}^{2}(9\Omega_{2}^{2}+2J_{1}\Omega_{2}+9J_{1}^{2})+2J(J_{1}+\Omega_{2})^{2}(2\Omega_{3}+J)}}\label{bb4}\\
&	\Lambda_{5}(t)=-\sqrt{5\Omega_{3}^{2}+2J\Omega_{3}+J^{2}+J_{1}^{2}+\Omega_{2}^{2}+\sqrt{2\Omega_{3}^{2}(9\Omega_{2}^{2}+2J_{1}\Omega_{2}+9J_{1}^{2})+2J(J_{1}+\Omega_{2})^{2}(2\Omega_{3}+J)}}
\label{bb3}	\\
&	\Lambda_{6}(t)=\sqrt{5\Omega_{3}^{2}+2J\Omega_{3}+J^{2}+J_{1}^{2}+\Omega_{2}^{2}-\sqrt{2\Omega_{3}^{2}(9\Omega_{2}^{2}+2J_{1}\Omega_{2}+9J_{1}^{2})+2J(J_{1}+\Omega_{2})^{2}(2\Omega_{3}+J)}}
\label{bb2}	\\
&	\Lambda_{7}(t)=-\sqrt{5\Omega_{3}^{2}+2J\Omega_{3}+J^{2}+J_{1}^{2}+\Omega_{2}^{2}-\sqrt{2\Omega_{3}^{2}(9\Omega_{2}^{2}+2J_{1}\Omega_{2}+9J_{1}^{2})+2J(J_{1}+\Omega_{2})^{2}(2\Omega_{3}+J)}}\label{bb1}
\end{align}
{and the corresponding eigenvectors can be expressed as}
\begin{align}
		&\ket{\Lambda_{0}(t)}=\dfrac{1}{2\sqrt{2}} \label{64}\\
		&\left(e^{-i\alpha(t)}\ket{\downarrow\downarrow\downarrow}+e^{-i\alpha(t)}\ket{\downarrow\downarrow\uparrow}+\ket{\downarrow\uparrow\downarrow}+\ket{\downarrow\uparrow\uparrow}+e^{-i\alpha(t)}\ket{\uparrow\downarrow\downarrow}+e^{-i\alpha(t)}\ket{\uparrow\downarrow\uparrow}+\ket{\uparrow\uparrow\downarrow}+\ket{\uparrow\uparrow\uparrow}\right) \nonumber\\
&		\ket{\Lambda_{1}(t)}=\dfrac{1}{2\sqrt{2}}\label{65}\\
		&\left(e^{i\alpha(t)}\ket{\downarrow\downarrow\downarrow} +\omega e^{i\alpha(t)}\ket{\downarrow\downarrow\uparrow}+i\ket{\downarrow\uparrow\downarrow}+i\omega\ket{\downarrow\uparrow\uparrow}-e^{i\alpha(t)}\ket{\uparrow\downarrow\downarrow}-\omega e^{i\alpha(t)}\ket{\uparrow\downarrow\uparrow}-i\ket{\uparrow\uparrow\downarrow}-i\omega\ket{\uparrow\uparrow\uparrow}\right) \nonumber\\
	&	\ket{\Lambda_{2}(t)}=\dfrac{1}{2\sqrt{2}}\label{66}\\
		&\left(e^{-i\alpha(t)}\ket{\downarrow\downarrow\downarrow} +ie^{-i\alpha(t)}\ket{\downarrow\downarrow\uparrow}-\ket{\downarrow\uparrow\downarrow}-i\ket{\downarrow\uparrow\uparrow}+e^{-i\alpha(t)}\ket{\uparrow\downarrow\downarrow}+ie^{-i\alpha(t)}\ket{\uparrow\downarrow\uparrow}-\ket{\uparrow\uparrow\downarrow}-i\ket{\uparrow\uparrow\uparrow}\right) \nonumber\\
	&	\ket{\Lambda_{3}(t)}=\dfrac{1}{2\sqrt{2}}\label{67}\\
		&\left(e^{i\alpha(t)}\ket{\downarrow\downarrow\downarrow} +i\omega e^{i\alpha(t)}\ket{\downarrow\downarrow\uparrow}-i\ket{\downarrow\uparrow\downarrow}+\omega\ket{\downarrow\uparrow\uparrow}-e^{i\alpha(t)}\ket{\uparrow\downarrow\downarrow}-i\omega e^{i\alpha(t)}\ket{\uparrow\downarrow\uparrow}+i\ket{\uparrow\uparrow\downarrow}-\omega\ket{\uparrow\uparrow\uparrow}\right) \nonumber\\
	&	\ket{\Lambda_{4}(t)}=\dfrac{1}{2\sqrt{2}}\label{68}\\
		&\left(e^{-i\alpha(t)}\ket{\downarrow\downarrow\downarrow} -e^{-i\alpha(t)}\ket{\downarrow\downarrow\uparrow}+\ket{\downarrow\uparrow\downarrow}-\ket{\downarrow\uparrow\uparrow}+e^{-i\alpha(t)}\ket{\uparrow\downarrow\downarrow}-e^{-i\alpha(t)}\ket{\uparrow\downarrow\uparrow}+\ket{\uparrow\uparrow\downarrow}-\ket{\uparrow\uparrow\uparrow}\right)\nonumber\\
	&	\ket{\Lambda_{5}(t)}=\dfrac{1}{2\sqrt{2}}\label{69}\\
		&\left(e^{i\alpha(t)}\ket{\downarrow\downarrow\downarrow}-\omega e^{i\alpha(t)}\ket{\downarrow\downarrow\uparrow}+i\ket{\downarrow\uparrow\downarrow}-i\omega\ket{\downarrow\uparrow\uparrow}-e^{i\alpha(t)}\ket{\uparrow\downarrow\downarrow}+\omega e^{i\alpha(t)}\ket{\uparrow\downarrow\uparrow}-i\ket{\uparrow\uparrow\downarrow}+i\omega\ket{\uparrow\uparrow\uparrow}\right)\nonumber
\end{align}		
	\begin{align}	
	&	\ket{\Lambda_{6}(t)}=\dfrac{1}{2\sqrt{2}}\label{70}\\
		&\left(e^{-i\alpha(t)}\ket{\downarrow\downarrow\downarrow}-ie^{-i\alpha(t)}\ket{\downarrow\downarrow\uparrow}-\ket{\downarrow\uparrow\downarrow}+i\ket{\downarrow\uparrow\uparrow}+e^{-i\alpha(t)}\ket{\uparrow\downarrow\downarrow}-ie^{-i\alpha(t)}\ket{\uparrow\downarrow\uparrow}-\ket{\uparrow\uparrow\downarrow}+i\ket{\uparrow\uparrow\uparrow}\right)\nonumber\\
		&\ket{\Lambda_{7}(t)}=\dfrac{1}{2\sqrt{2}}\label{71}\\
		&\left(e^{i\alpha(t)}\ket{\downarrow\downarrow\downarrow}-i\omega e^{i\alpha(t)}\ket{\downarrow\downarrow\uparrow}-i\ket{\downarrow\uparrow\downarrow}-\omega \ket{\downarrow\uparrow\uparrow}-e^{i\alpha(t)}\ket{\uparrow\downarrow\downarrow}+i\omega e^{i\alpha(t)}\ket{\uparrow\downarrow\uparrow}+i\ket{\uparrow\uparrow\downarrow}+\omega\ket{\uparrow\uparrow\uparrow}\right)\nonumber
\end{align}
where we have defined 
\begin{align}
&\alpha(t) = \frac{\pi}{4}-\kappa(t)\\
& \tan[\kappa(t)]=\frac{\Omega_{2}(t)}{2J_{1}}+\frac{\Omega_{3}(t)}{2J}.
\end{align}

\end{appendices}

\end{document}